\documentclass{aa}  

\usepackage{graphicx}
\usepackage{txfonts}
\usepackage{xcolor}
\usepackage{tabularx}
\usepackage{appendix,natbib}
\usepackage{amsmath}
\usepackage{subfigure}
\usepackage[breaklinks=true]{hyperref} 

\usepackage{pgfplotstable} 
\usepackage{csvsimple}
\usepackage{lscape} 

\def\hz{high-$z \ $}

\newcommand{\fdm}{{f_{\textrm{\tiny DM}}}}
\newcommand{\Ms}{\mathrm{M_\odot}}

\DeclareUnicodeCharacter{2212}{-}

\begin{document}

   \title{Observational evidence of evolving dark matter profiles at $z\leq 1$}


   \author{Gauri Sharma, 
          \inst{1,2,3}\fnmsep\thanks{Contact: gsharma@sissa.it}
          Paolo Salucci \inst{1,2,3}
          \and
         Glenn van de Ven\inst{4}
          }

   \institute{SISSA International School for Advanced Studies, Via Bonomea 265, 34136 Trieste, Italy
         \and
           GSKY, INFN-Sezione di Trieste, via Valerio 2, 34127 Trieste, Italy
         \and
           IFPU Institute for Fundamental Physics of the Universe, Via Beirut, 2, 34151 Trieste, Italy
         \and 
           Department of Astrophysics, University of Vienna, T\"urkenschanzstrasse 17, 1180 Wien, Austria               
             }

   \date{Received XXXX; accepted XXXX}

 
  \abstract
   {In the concordance cosmological scenario, the cold collisionless  dark matter component dominates the mass budget of galaxies and interacts with baryons only via gravity. However, there is growing evidence that the former, instead,  responds to the baryonic (feedback) processes by modifying its density distribution. These processes can be captured by comparing the inner dynamics of galaxies across cosmic time.}
   {We present a pilot study of dynamical mass modeling of high redshift galaxy rotation curves, which is capable of constraining the structure of dark matter halos across cosmic time.}
   {We investigate the dark matter halos of 256 star-forming disk-like galaxies at $z\sim 1$ using the KMOS Redshift One Spectroscopic Survey (KROSS). This sample covers the redshifts $0.6 \leq z \leq 1.04$, effective radii $0.69 \leq R_e [\mathrm{kpc}] \leq 7.76$, and total stellar masses $8.7 \leq log(M_{\mathrm{star}} \ [\mathrm{M_\odot}]) \leq 11.32$. We present a mass modeling approach to study the rotation curves of these galaxies, which allow us to dynamically calculate the physical properties associated with the baryons and the dark matter halo. For the former we assume a Freeman disk, while for the latter we employ the NFW (cusp) and the Burkert (cored) halo profiles, separately. At the end, we compare the results of both cases with state-of-the-art galaxy simulations (EAGLE, TNG100, and TNG50).}
   {We find that the "cored" dark matter halo emerged as the  dominant quantity from  a radius 1-3 times the effective radius. Its fraction to the total mass is in good agreement with the outcome of hydrodynamical galaxy simulations. Remarkably, we found  that the dark matter core of $z\sim 1$ star-forming galaxies are smaller and denser than their local counterparts. }
   {Dark matter halos have gradually expanded over the past 6.5 Gyrs. That is, observations are capable of capturing the dark matter response to the baryonic processes, thus giving us the first piece of empirical evidence of "gravitational potential fluctuations" in the inner region of galaxies that can be verified with deep surveys and future missions. \thanks{Data availability: For physical properties of galaxies, such as stellar mass, gas mass, SFR, velocity, we refer readers to the catalogs released with \citealt[][]{GS21a} (Flat RCs catalog \citealt{FlatRCs_GS21a}) and \citealt{GS21b} (DM fraction catalog \citealt{DMfraction_GS21b}). The measurements from coadded RC mass modeling are available in form of tables in Appendix~\ref{sec:table} and measurements from individual RCs are attached in the external appendix, which will be released with the publication.  }}


   \maketitle
%

\section{Introduction}
\label{sec:Intro}
Dark matter (DM) is a type of matter that astrophysicists have brought forward to account for the observational effects that appear to be due to the presence of an invisible massive component \citep[][and references therein]{Opick1915, Kapteyn1922, Jeans1922, Oort1932, Zwicky1933, Smith1936, Roberts1973, Rubin1978, bosma1981, Swart2017}. The astrophysical nature of such invisible matter can be inferred from its gravitational interaction on luminous matter \citep[][references therein]{Ostriker1973, PS1988, PS1996, NFW1997,  Sofue2001, PS2019} and on radiation which emerges at different cosmological scales \citep[][references therein]{Sunyaev1969, Silk1993, Boehm2005, Chluba2011, Cora2014, Slatyer2018, Ali2019}. On the other hand, its particle nature, which is the center of the DM phenomenon, has not been characterized yet. 

Recently, the DM phenomenon was uniquely framed into the dark energy driven cold dark matter ($\mathrm{\Lambda CDM}$) scenario \citep{Turner1985, Turner1990, Ellis2018}, where nonrelativistic DM can be described as a collisionless fluid, whose particle with a cross-section of about $10^{-26} \ \mathrm{cm^{2}}$, only interacts gravitationally and very weakly with standard model particles. This scenario, which emerges very clearly cosmologically \citep{Peeblesbook, Coles2002}, has been confirmed by large-scale observations \citep[][and references therein]{Einasto2017, Lorenzo2018} and can be investigated by appropriate simulations \citep[][and references therein]{Klypin1996, Vogelsberger2020} and semi-analytical modeling \citep[][and references therein]{Kauffmann1993, Rachel1999, Lacey2001, Guo2011, Rachel2015}. As a consequence, a very well-defined perturbation power law spectrum and lack of dependencies on the initial conditions, yield to a specific bottom-up formation and evolution of DM halos. However, two aspects in the progressive field of DM opposes the latter: 1) the search of DM particles, which is performed by various direct and indirect methods \citep[e.g.,][]{Covi2013, Gaskins2016} and one of the major efforts of astro-particle physics, has been unsuccessful so far \citep{Arcadi2018}; 2) the main imprint of a collisionless particle is a  DM halo with cuspy central DM density ($\rho \propto r^{-1}$), which is not present in the majority of local spirals, dwarf disks, and low surface brightness galaxies \citep[e.g.][]{palunas2000, PS2000, Block2001, Karukes2017, Chiara2019, salet}. Moreover, local galaxies show that the physical properties (e.g., luminosity and disk scale length) of luminous matter are deeply connected with the structural properties of DM (e.g., its core radius and core density); readers can refer to \citet[][]{PS2019} and \citet{Pengfei2020}. This is very difficult to reconcile with collisionless particle, taking into account that most of the DM resides inside the galactic halo. These issues imply that in order to tackle the DM phenomenon, we need to refine the adopted DM scenario. 

At this point, it is important to point out that besides the structure of present-day DM halos, the DM halo evolution through cosmic time may also reflect the nature of a DM particle: in the $\mathrm{\Lambda CDM}$ scenario, the collisionless nature of a dark particle is itself imprinted in the dark halo evolution with time \citep{Ellis2018}, which is well followed by hydrodynamical simulations \citep{Jian2020}. Also in the other scenarios of DM such as warm DM, boson condensates, interacting DM, ultra light axions, etc., and with alternative models of DM such as MOdified Newtonian Dynamics (MOND), one can find correspondence between properties of DM halos around local objects and those at high redshift (hereafter high-$z$). 

In the same vein, we must remind readers that, until recently, the  determination of \hz DM halo density profiles was impossible: the kinematical measurements were lacking the sufficient  sensitivity and spatial resolution. But, in the last decade, advanced use of integral field units (IFUs) in galaxy surveys has opened up possibilities for studying the spatially resolved kinematics of \hz galaxies, for example, surveys with the Multi-Unit Spectroscopic Explorer (MUSE: \citealt{MUSE}), K-band Multi-Object Spectrograph (KMOS: \citealt{sharples_2014}), the Spectrograph for INtegral Field Observations in the Near Infrared (SINFONI: \citealt{SINFONI}), and Atacama Large Millimeter/submillimeter Array (ALMA: \citealt{cox2016}). A detailed review by \citet[][references therein]{Forster2020} highlights the main findings of the aforementioned surveys. In the context of dynamical studies, \citet{Forster2020} show that in using the resolved kinematics of galaxies at high-$z$, we can now obtain the rotation curves (RCs) of galaxies up to $z\approx 1-2.5$. These RCs allow us to probe the baryon and DM content on galactic scales, as well as their distribution. In particular, \citet{Genzel2020} dynamically mass modeled the RCs of 41 star-forming galaxies (SFGs), and provided the first piece of evidence of a cored DM profile at high-$z$ (in the majority of the sample). Moreover, in a more recent study, \citet{Bouche2021} dynamically mass-modeled the nine $z\approx 1$ SFGs and they show that, similar to the $z\sim 1$ sample of \citet{Genzel2020}, a significant fraction of their sample also shows evidence of a cored DM profile. The study of \citet{Bouche2021} also shows that the DM "cored" feature is generally associated with galaxies with low stellar-to-halo mass, while "cusps" are found to be a function of strong star-formation activity. They conclude that this suggests feedback-induced core formation in the $\Lambda$CDM scenario. Now, in order to extend our understanding of high-$z$ DM halos, in particular the cusp-core transformation, and to develop accurate models of galaxy evolution, we in this work focus on detailed mass modeling of the largest set of the currently available $z\sim 1$ rotation curve of SFGs published and discussed  in our previous works \citep{GS21a, GS21b}.

Briefly, in \citet{GS21a}, we used KMOS Redshift One Spectroscopic Survey (KROSS) data to derive the kinematics of a large sample of star-forming disk-like galaxies at $z\sim 1$. We converted these observed kinematics into intrinsic rotation curves by correcting the issues related to \hz measurements, such as, beam-smearing\footnote{\tiny{Without Adaptive Optics (AO), an IFU achieves only $0.5\arcsec -1.0 \arcsec$ spatial resolution, whereas, a galaxy from $z\sim 1$ has a typical angular size of $2\arcsec - 3\arcsec$. The finite beam size causes the line emission to smear on the adjacent pixels. This effect is referred to as `Beam Smearing', which under estimates the rotation velocity and overestimates the velocity dispersion.}} and pressure-support\footnote{\tiny{the kinematics of the galaxies are derived using the emission lines, such as H$_\alpha$ or [OIII]. These emission lines arise from the gaseous disk around the stars or the inter-stellar medium (ISM). If the ISM is highly turbulent, then the emission also experiences a turbulence, i.e., radial force against gravity. This turbulence or force scales with the gas density and velocity dispersion. Since the density and velocity dispersion both decrease with increasing radius, this creates a pressure gradient, i.e., $F_{P} \propto -dP/dr$ (where $P \propto \rho \sigma^2$). The resulting radial force supports the disk and makes it rotate slower than the actual circular velocity, which might result in declining the RCs and potentially underestimate the dynamical masses}} conditions. In particular, we used 3D-forward modeling of datacubes using $^{3D}$Barolo \citep{ETD15}, which provides somewhat better constraints on kinematics than 2D-kinematic modeling approaches. For pressure support, we employed the pressure gradient correction (PGC) method, which makes use of all of the information available in datacube and therefore provides better results than pressure support corrections that assume isotropic and constant velocity dispersion; for details we refer the interested reader to \citet[][appendix]{GS21a}. This allows us to have a large sample of more than 200 accurate RCs of disk-like galaxies at $z\sim 1$ (i.e., 6.5 Gyr lookback time). This large amount of data can help us open a new  portal to the nature of DM, as well as allow us to test the $\Lambda$CDM collisionless scenario.  

In the follow-up study, \citet{GS21b}, we employed a halo model independent approach of DM determination using the RCs established in \citet{GS21a}, where the  stellar mass ($M_*$) of the objects were estimated by using mass-to-light ratios derived from fitting the spectral energy distribution of the galaxies, the total gas (molecular and atomic) mass by means of  the scaling relations \citep{Tacconi2018, Lagos2011}, and the dynamical mass ($M_{dyn} = G^{-1} V^2 \ R$)  from the rotation curves, evaluated at a different scale radius (disk radius: $0.59R_e$, optical radius: $1.89R_e$, and outer radius: $2.95R_e$ ). These measurements provide us with DM fractions ($f_{DM} = 1 - M_{bar}/M_{dyn}$) at different radii. This work shows that the galaxies are DM dominated from $1-3 R_e$. The majority ($>72\%$) of star-forming galaxies in our sample have outer disks ($\sim 5−10$ kpc) dominated by DM, which agrees well with local SFGs.

In the present work, we investigate the kinematics of our same star-forming disk-like galaxies from $z\sim 1$ and dynamically mass-model their RCs. The latter is a model-dependent approach for deriving the physical quantities such as the DM fraction, stellar masses, and effective radii. Thus, comparing the two studies, \citet{GS21b} and the current work allowed us to check the robustness of our dynamical models. Through the mass-modeling techniques, we also determined the structural properties of DM halos and checked the scaling relations between dark and baryonic matter, which are observed in local star-forming disk galaxies. In the end, we compared our work with the recent study of \citet{Genzel2020}, in which the authors dynamically mass model the galaxy kinematics, which was derived from the 1D software Slit Extraction Method. The latter method is known to bias the kinematic extraction (i.e., RCs), as shown in \citet{ETD15} and \citet{ETD16}. However, a complementary study of the work by \citet{Genzel2020} was done by \citet{Price2021} using 3D-forward kinematic models; however, they obtained the same results as \citet{Genzel2020}. Besides this, both studies corrected the pressure support under the assumption of isotropic and constant velocity dispersion using \citet{Burkert2010} method, which underestimates the rotational velocity, as shown in \citet{GS21a}. Despite the significant discrepancies between our work and the aforementioned studies, we compared them in terms of the DM fraction and its structural properties.

This article is organized as follows: Section~\ref{sec:data} contains a brief discussion on the data employed. In Section~\ref{sec:method} we provide the mass models to disentangle the baryons and DM component from RCs and describe their fitting techniques. In Section~\ref{sec:results} we show the results and critically analyze them. In Section~\ref{sec:sims-comp} we compare our results with current state-of-the-art simulations. Finally in Section~\ref{sec:discussion} and Section~\ref{sec:summary}, we discuss  the results and provide a summary with future prospectives. Throughout the analysis we assumed a flat $\Lambda$CDM cosmology with $\Omega_{m,0} =0.27$, $\Omega_{\Lambda,0}=0.73$, and $H_0=70 \ \mathrm{km \ s^{-1}}$.


\section{Data}
\label{sec:data}
We studied the KROSS parent sample previously studied by \citet{stott2016} and others \citep{H17, AT2019, AT2019b, HLJ17}. This sample contains 586 SFGs from  $z\sim 1$.  We selected 344 objects out of 586, on the basis of an integrated $H\alpha$ flux cut ($F_{H\alpha}>2\times 10^ {-17} \  \mathrm{[erg \ s^ {-1} \ cm^ {-2}]} $) and inclination angle ($25^{\circ} \leq \theta_i \leq 75^{\circ} $). The chosen flux and inclination cuts ensure sufficient signal-to-noise ratio (S/N) data and reduce the impact of extinction\footnote{In the highly inclined system ($\theta_i>75^{\circ}$) observed flux is extinct due to extinction, which suppresses the rotation velocity up to a few times $R_e$ \citep[see][]{Valotto2003}. On the other hand, in face-on galaxies ($\theta_i<25^{\circ}$) the rotation signal drops below the observational uncertainties. Therefore, to be conservative we down-selected the sample for $25^{\circ} \leq \theta_i \leq 75^{\circ} $.} during the kinematic modeling.  The intrinsic characteristic of the selected sample is the following (given with respect to TableA1 of \citealt[][hereafter H17]{H17}): 1)the AGN-flag is zero, that is there no evidence for an AGN contribution to the $H_\alpha$ emission-line profile; and 2)quality flag of objects is 1, 2, and 3, that is only $H_\alpha$-emission line detected objects ($S/N>3$) are selected. We adopted the values of requisite quantities, such as effective radii ($R_e$), stellar mass ($M_*$), and redshift ($z$) from \citet{H17} and \citet{stott2016}. Moreover, the reddening corrected $H_\alpha$ star-formation rate ($SFR_{H\alpha}$) of the sample were adopted from \citet{GS21b}. We remark that the parent sample was selected in such a way that it is biased against galaxies in a merging or interacting state, and thus should provide a representatives sample of main sequence SFGs at $z\sim 1$ \citep[see][]{stott2016}.

As briefly mentioned above, the stellar masses and effective radii are two relevant physical properties of galaxies that we adopted from H17 and used throughout this work. In H17,  stellar masses were calculated using a fixed mass-to-light ratio following $M_* = \Upsilon_H \times 10^{-0.4\times(M_H - 4.71)}$, where $\Upsilon_H$ and $M_H$ are the mass-to-light ratio and absolute magnitude in the H band (rest frame), respectively.  Here $\Upsilon_H=0.2$, which is the median value of our sample obtained using the {\sc Hyperz} \citep{Bolzonella2000} spectral energy distribution (SED) fitting tool. These stellar masses are in good agreement with those inferred using a full SED-fitting approach \citep{AT2019}, with no noticeable offset and only a 0.2 dex uncertainty \citep[see][appendix]{GS21b}. The effective radii of the sample were measured from broadband images by deconvolving the point spread function (PSF) and semi-major axis of the aperture, which contains half of the total flux of a galaxy; for details, readers can refer to \citet{H17}.

\subsection{Individual RCs} \label{sec:IRCs}
To extract kinematics of a selected sample (344 SFGs), we exploited publicly available datacubes (data availability\footnote{KROSS: http://astro.dur.ac.uk/KROSS/data.html}), and the kinematics of these objects were re-derived from KROSS datacubes employing the 3D-BAROLO code \citep{ETD15, ETD16}. The latter takes into account the beam smearing correction in 3D-space and provides moment maps, the stellar surface brightness profile, the RC, the dispersion curve (DC), along with the kinematic models. After running the 3D-Barolo, we assessed the quality of the objects, which is defined by taking the S/N into account, and limitation of 3D-Barolo (e.g., inclination). The best quality objects, that is to say quality-1 and 2, are the ones that have a $\mathrm{S/N}\geq3$\footnote{Here S/N corresponds to the signal in each spaxel in the masked region of a datacube.} and an inclination $30^{\circ} \leq \theta_i \leq 75^{\circ}$.  Apart from this, our RCs were corrected for pressure gradients, which likely affect the kinematics of high-$z$ galaxies. Out of the 344 KROSS objects, we chose to analyze only 256 rotation dominated objects, referred to as the Q12 sample, for details we refer the reader to \citet{GS21a}. In the following work, we use this Q12 sample to analyze the DM distribution.

\subsection{Coadded RCs} \label{sec:CRCs}
\begin{figure*}
  \begin{center}
   \includegraphics[angle=0,height=11.0truecm,width=18.5truecm]{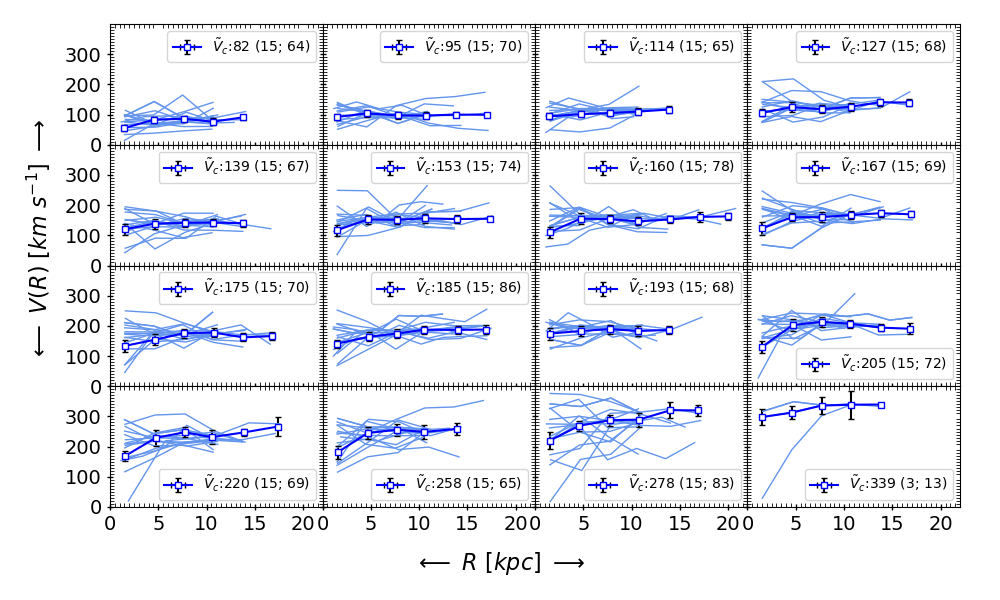}       
    
    \caption{Coadded and binned rotation curves.  The individual and binned RCs are shown by blue thin curves and squares connected with a dark blue line, respectively. The central velocity ($\tilde{V}_c$) and the number of data points per bin are printed in the legend of each panel. For example, the notation $\tilde{V}_c:82$ (15; 64) means that the central circular velocity of the bin is 82 $\mathrm{km \ s^{-1}}$ and that it contains 15 RCs and a total of 64 data-points. We note that here $\tilde{V}_c$ is the circular velocity computed at $\tilde{R}_{out}$ ($=2.95 \ R_e$).}
    \label{fig:coadded-RCs}
  \end{center}
\end{figure*}

In this work, we will investigate the individual RCs of the Q12 sample, as well as their coadded RCs (or stacked RCs). The process of coaddition, which we briefly recall here, is accurately described in \citet{PS1996} and \citet{PS2019}. That is, the individual RCs were coadded according to their circular velocities $V_c$ (computed at $R_{out}=2.95 \ R_e$), keeping 15 RCs per bin, except for the highest velocity bin. Then, we binned the galaxies radially per $2.5 \ kpc$ corresponding to the binning scale of the 3D-Barolo; for details, readers can refer to \citet{GS21a}. For the binning, we used the standard weighted mean statistic given by $\bar{X} = \frac{\Sigma_{i=1}^{n}x_i \times w_i}{\Sigma_{i=1}^{n}w_i}$, where, $ x_i=\mathrm{data}, \ w_i=1/error^2, \ \mathrm{and} \ \bar{X}=\mathrm{binned \ data}$. The errors on the binned data are the root mean square error, computed as $\delta_{v}^{r_{i}} = \sqrt{\sum([(\delta_{v_{i}}^r)^2 + (\sigma_{v}^{r_{i}})^2])/N}$, where $\delta_{v_{i}}^r$ is the individual error on the velocities per radial bin and $\sigma_v^{r_{i}}$ is scatter per radial bin. This process yields 16 coadded RCs, each one has a minimum of five independent data-points (see Figure~\ref{fig:coadded-RCs}, and for more details on the binning process we refer the readers to \citet{GS21a}). While binning the RCs, we also coadded and binned the physical quantities (e.g., $M_{\mathrm{D}}, R_{\mathrm{D}}$) associated with individual objects per velocity bin using RMS statistics and the errors are standard errors. These binned parameters were used in RC modeling either as an initial guess (when the parameters have a flat prior) or as a central value in the case of a Gaussian prior. In the following text, the binned parameters are represented by a tilde character on top of them (e.g., $\tilde{M}_{\mathrm{D}}$) and coadded plus binned RCs are referred to as "coadded RCs".

We remark that the main advantage of the coadded RC is that it allows for the statistical investigation of a particular class of objects, especially when the data quality is low and the number of observations are limited. The latter is currently the case for \hz galaxies. Moreover, $H_\alpha$ emission is the only way to trace the kinematics of \hz galaxies in the optical (and near-infrared) wavelength range. At the same time we know that $H_\alpha$ originates from star-forming regions, that is the interstellar medium, which is turbulent at high-$z$; therefore, the individual RCs are generally noisy, that is to say artificially wiggled \citep{PhysicsOf-ISM1980}. In these mixed situations with poor data quality and limited observations, the coaddition of RCs acts as a rescue tool, providing a smooth circular velocity of the galaxies and reducing minor observational errors in the data. However, the coaddition of RCs also has some drawbacks, for example, the following can not be evident: 1) the diversity in the shape of the RCs of a given class of galaxies and 2) the information on noncircular motions due to a bar- and bulge-like component. The former and latter both are interlinked and arise due to a dominant triaxial halo which induces bi-symmetric flows in the gas component of a galaxy as reported in \citet[][]{Marasco2018} and \citet{Kyle2019}. Despite these aspects, and taking the current situation of \hz observations into account where we have a lack of data and no resolution within 4 kpc, we propose to study coadded as well as individuals RCs.

Furthermore, before performing the coaddition of the RCs, we visually inspected each individual RC, and verified that the individual RCs of each velocity bin have more or less the same radial profile. Quantitatively, the data points of each velocity bin have a maximum dispersion of $\sim0.1$ dex, which is very similar to the study performed by \citet{PS1996} for local SFGs. Thus, our method of coaddition is applicable and certainly necessary for \hz data, where we have moderate signal-to-noise, low resolution, and limited observations. An alternative way to avoid the abovementioned problems of \hz observations is "smooth RCs," where we fit the observed RC with a parametric form, for example arctan ($V_{rot} \propto \arctan(R)$), then this best-fit RC can be used for mass modeling or coaddition. The main advantage of smoothed RCs is that while mass-modeling with Markov
chain Monte Carlo (MCMC), they do not face the problem of nonconvergence.
However, we have two important reasons not to use smoothed RCs. First, observed RCs are difficult to fit when they are wiggled (or declining), and thus they cannot be smoothed with simple parametric forms of RCs. Second, a statistical study of RCs, that is coaddition and binning performed in our work, leads to averaging RCs, which is similar to smoothing. Therefore, it is not advisable to smooth the RCs a priori, otherwise, twice the information may be lost. Moreover, our study focuses on the shape of the RCs \citep{GS21a}, the DM fraction \citep{GS21b}, the component separation (current work), and the derivation of the structural properties of the DM. These studies are very sensitive to the intrinsic shapes of the RCs. Therefore, we did not risk smoothing the RCs at outset but used the observed RCs.


\section{Methodology}
\label{sec:method}
The RC of a galaxy gives us the total circular velocity of the system as a function of its radius. This $V_c(R)$ is composed of the contribution of baryons (stars and gas) and DM. Therefore, $V_c(R)$ can be described as a sum in quadrature of the following:
\begin{equation}
\label{eq:vtot}
V^2_c(R) = V^2_{\mathrm{D}}(R) + V^2_{\mathrm{bulge}}(R) + V^2_{\mathrm{H2}}(R) + V^2_{\mathrm{HI}}(R) + V^2_{\mathrm{DM}}(R) ,
\end{equation}
where $V_{\mathrm{D}}(R), V_{\mathrm{bulge}}(R) , V_{\mathrm{H2}}(R)$, and $V_{\mathrm{HI}}(R)$ are the circular velocity profiles of the stellar disk, stellar bulge, molecular (H2) and atomic (HI) gas disks, together providing the baryon's contribution to the circular velocity and $V_{\mathrm{DM}}(R)$ gives the DM contribution. In the following sections, we present an approach to disentangle the aforementioned contributions from observed RCs. We note that in \citet{GS21a}, we corrected the observed RCs for inclination and pressure support, that is RCs are very close to intrinsic circular velocity curves.

\subsection{Mass models}
\label{sec:models}	
Here, we explain the various components of our mass modeling.

\subsubsection{Stellar and gas disk} The stellar and gaseous components of local star-forming disks, dwarf-disks and low surface brightness galaxies are well described by the exponential disk model of \cite{Freeman1970} \citep[e.g.,][] {PS1996, Karukes2017, Chiara2019}. Here, we assume that the distribution of stars and gas in $z\sim 1$ star-forming disk galaxies are also in exponential disks, so that the surface densities are as follows:
\begin{equation}
\label{eq:freeman-density}
\Sigma_{\_}(R) \propto \frac{M_{\_}}{R_{scale}} \exp(-R/R_{scale}),
\end{equation}
where $M_{\_}$ and $R_{scale}$ are the total mass and the scale length of the different components (stars, H2, and HI), respectively. We note that stellar mass in the disk (without a bulge) is denoted by $M_{\mathrm{D}}$, while the contribution of the total stellar mass is represented by $M_{\mathrm{star}} = M_{\mathrm{D}} + M_{\mathrm{bulge}}$. Given the density distribution of the stars and the gas, their contribution to the circular velocity of the disk can be expressed as follows:
\begin{equation}
\label{eq:Vd}
V_{\_}^2(R)= \frac{1}{2} \Big( \frac{GM_{\_}}{R_{scale}} \Big) \ (x^2) \  [I_0K_0 - I_1K_1],
\end{equation}
where, $x = R/R_{scale}$ and $I_n$ and $K_n$ are modified Bessel functions computed at $1.6x$ for stars and $0.53x$ for gas \citep[c.f.][]{PS1996, Karukes2017}.
                  
\subsubsection{Bulge component} First, we would like to emphasize that our RCs are not well resolved within the 5 kpc where we expect the bulge to contribute significantly, that is, the bulge is unresolved in our data. Thus, we assume that the bulge is a point mass, so that its contribution to the circular velocity becomes
\begin{equation}
\label{eq:Vb}
V_{\mathrm{bulge}}^2(R)= \frac{G\ M_{\mathrm{bulge}}}{R},
\end{equation}
where, $M_{\mathrm{bulge}}$ is the total bulge mass, and it can be estimated using bulge-to-total ratio $B/T = M_{\mathrm{bulge}}/M_{\mathrm{star}}$. 


\subsubsection{DM halo component} 
\label{sec:DM-profile}
In this work, we test two widely used DM halo models, namely the profiles of Burkert \citep{Burkert} and NFW \citep{NFW}.
\\ \\
\underline{Burkert Halo:} Local spirals and low surface brightness galaxies suggest existence of central DM cores, that is $\rho_{inner} \propto const$ \citep{PS2000, Block2001}. This observed DM-profile is well fitted by the \citet{Burkert} halo, which possesses a double power law in the DM-density, that is, at small radii $\rho \propto R^{0}$ and at larger radii $\rho \propto R^{-3}$. Such a DM-density distribution can be expressed as follows: 
\begin{equation}
\label{eq:BurkertDensity}
\rho (R) = \frac{\rho_{_0}}{ \Big(1+ \frac{R}{r_{_0}} \Big) \Big( 1+ \frac{R^2}{r_{_0}^2}  \Big)   ,      }
\end{equation}
where $\rho_{_0}$ and $r_{_0}$ are the central DM core density and core radius, respectively. Assuming spherical symmetry, the mass profile of the Burkert DM halo can be expressed as follows: 
\begin{equation}
\label{eq:BurkertMass}
M_{\mathrm{DM}}^{\mathrm{Burk}} (R) = 4 \pi \rho_{_0} r_{_0}^3 \Big[ \ln \Big(1+\frac{R}{r_{_0}} \Big) - \arctan \Big(\frac{R}{r_{_0}} \Big) + 0.5 \ \ln \Big(1+\frac{R^2}{r_{_0}^2} \Big)  \Big].
\end{equation}
%
%
\underline{NFW Halo:} In the standard $\Lambda$CDM paradigm, the current cosmological simulations predict a cuspy DM distribution in the center, that is, $\rho_{inner} \propto R^{-1}$. This type of DM profile is well approximated by the NFW halo, which is again a double power law; however, it has $\rho \propto R^{-1}$ at small radii and $\rho \propto R^{-3}$ at larger radii. Such a DM-density distribution can be expressed as: 
\begin{equation}
\label{eq:NFWDensity}
\rho (R)  = \frac{\rho_{_s} }{\Big(\frac{R}{r_{_s}} \Big) \ \Big( 1+\frac{R}{r_{_s}} \Big)^2} ,
\end{equation}
where $\rho_{_s}$ and $r_{_s}$ are the characteristic density and scale radius of the DM distribution, respectively. Assuming spherical symmetry, the mass profile of the NFW DM halo is as follows:
\begin{equation}
\label{eq:NFWMass}
M_{\mathrm{DM}}^{\mathrm{NFW}} (R) = 4\pi \rho_{_s} r_{_s}^3 \Big[\ln \Big(1+\frac{R}{r_{_s}} \Big) - \frac{\frac{R}{r_{_s}}}{1+ \frac{R}{r_{_s}}}   \Big].
\end{equation}
The circular velocity of the above DM profiles can be written as: 
\begin{equation}
\label{eq:Vhalo}
V_{\mathrm{DM}}^2 (R) =  G \frac {\ M_{\mathrm{DM}} (<R)}{R},
\end{equation}
where $M_{\mathrm{DM}} (<R)$ is the enclosed DM mass within radius $R$ given in Equation~\ref{eq:BurkertMass} \& \ref{eq:NFWMass}. The above formalism allowed us to compute the various structural properties of the DM halo, namely $r_{_{0/s}}$ and $\rho_{_{0/s}}$.\footnote{In $\rho_{_{0/s}}$ and $r_{_{0/s}}$, subscript $0$ gives the core radius and core density in the case of the Burkert halo, while subscript $s$ defines the scale radius and characteristic density of the NFW halo.} For the purpose of this work, we also define the virial radius ($R_{\mathrm{vir}}$), virial mass ($M_{\mathrm{vir}}$), and concentration ($c_{vir}$) parameter of the halo. The virial radius is the distance within which the average halo density is 200 times the critical density ($\rho_{\mathrm{crit}} = 9.3 \times 10^{-30} gm \ cm^{-3}$) of the universe, it can be easily computed numerically. The virial mass is defined as follows:
\begin{equation}
\label{eq:Mvir}
M_{\mathrm{vir}} = \frac{4}{3} \pi \ R_{\mathrm{vir}}^3  \  200 \  \rho_{\mathrm{crit}}.
\end{equation}
In the case of the NFW halo, the ratio of the virial radius to scale radius gives the halo concentration, $c_{vir}=R_{\mathrm{vir}}/r_{_s}$. Furthermore, once the various components of the RC are disentangled, then, given the information on $V_{\mathrm{DM}}(R)$ and $V_{c}(R)$, one can also estimate the following:
\begin{equation}
\label{eq:fdm}
\fdm (<R) = \frac{V_{\mathrm{DM}}^2(R)}{V_{c}^2(R)},
\end{equation} 
which gives us the DM fraction within the radius $R$. 

\begin{figure*}
 \begin{center}
  \includegraphics[angle=0,height=6.6truecm,width=8.0truecm]{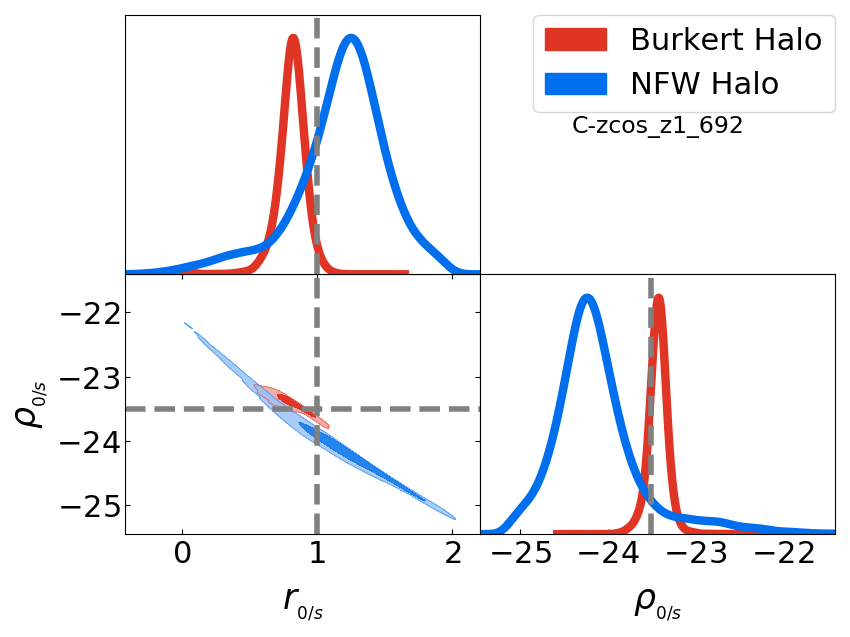} 
   \includegraphics[angle=0,height=6.5truecm,width=8.0truecm]{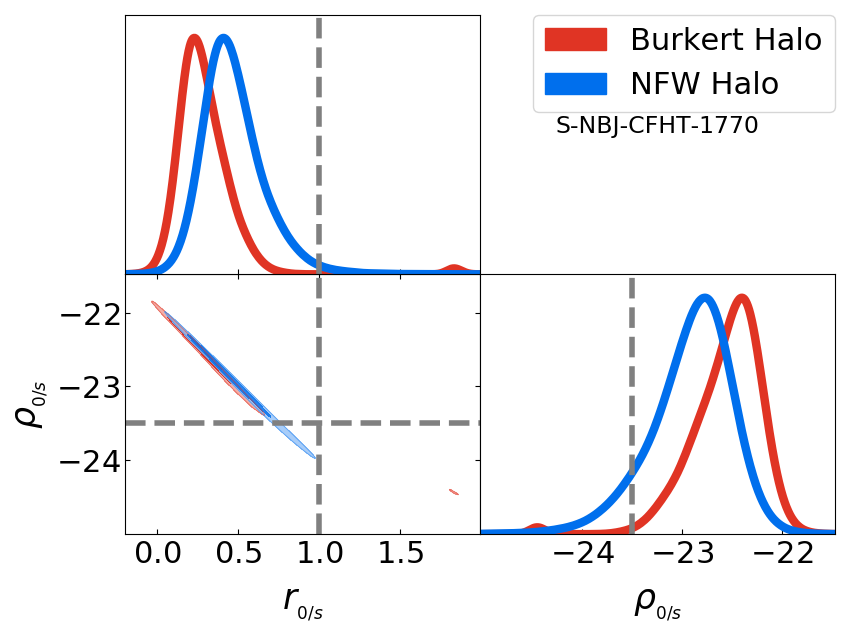}  
  
  \includegraphics[angle=0,height=7.0truecm,width=8.5truecm]{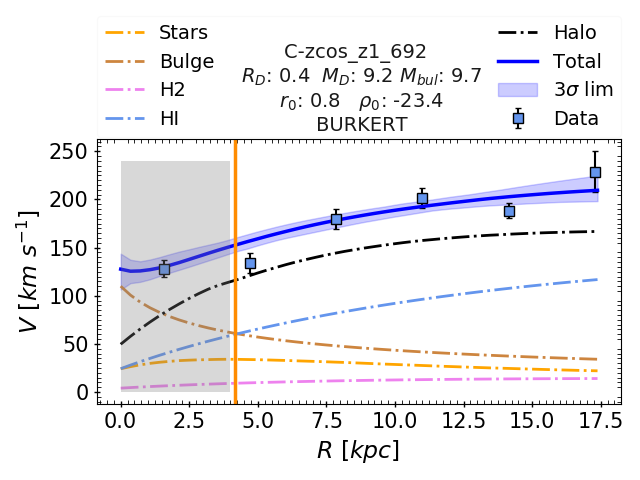}  
  \includegraphics[angle=0,height=7.0truecm,width=8.5truecm]{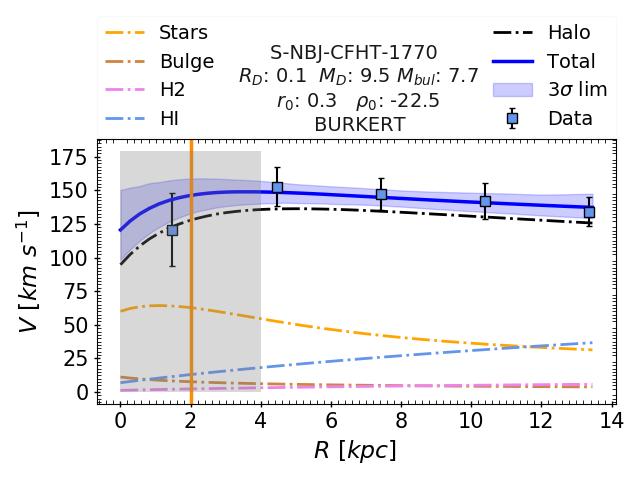}    
  
  \includegraphics[angle=0,height=7.0truecm,width=8.5truecm]{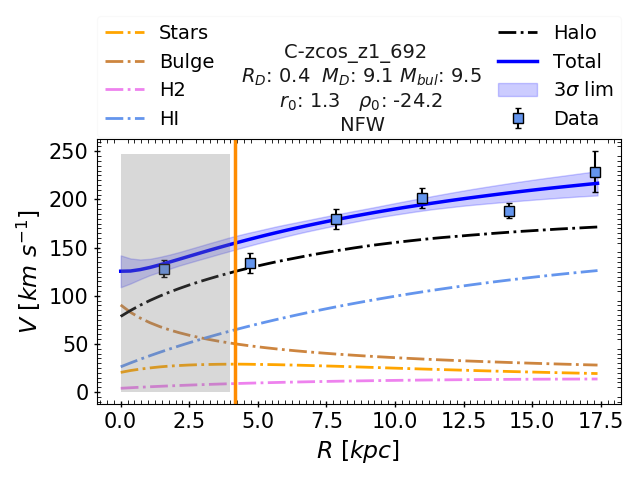}  
  \includegraphics[angle=0,height=7.0truecm,width=8.5truecm]{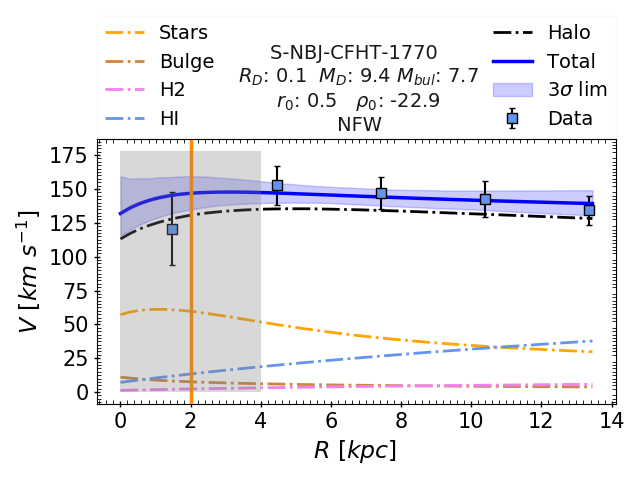}

    \caption{\underline{Individual RC mass modeling}. {\em Row 1:} Posterior distributions (MCMC output) of the following estimated parameters: $r_{_{0}}$ and $\rho_{_{0}}$ for Burkert (red) and $r_{_{s}}$ and $\rho_{_{s}}$ for the NFW (blue) halo. The vertical and horizontal gray dashed lines show the initial guess for parameters $r_{_{0/s}}$ and $\rho_{_{0/s}}$. The full posterior distribution of parameter space $\vec{\theta}$ of these example cases are shown in Figure~\ref{fig:IRCs_full}. {\em Rows 2 \& 3:} Best fit to the RCs and their velocity decomposition for Burkert and NFW halos, respectively. In the RC panels, the blue square with error bars represents the observed data, the blue solid line shows the best fit to the data, and the blue shaded area represents the $3\sigma$ error in the fit. The decomposed velocity of stars, molecular gas, atomic gas, bulge, and dark matter is represented by yellow, violet, light blue, brown, and black dotted-dashed lines, respectively. The vertical orange line shows the effective radius and the gray shaded area represents the unresolved region of the RC. The best-fit value of parameter space $\vec{\theta}$ is given in the legend of the figures. All the parameters are printed in log, having physical units as follows: $M_{\mathrm{D}} \ [\mathrm{M_\odot}]$, $R_{\mathrm{D}} \ [\mathrm{kpc}]$, $M_{\mathrm{bulge}} \ [\mathrm{M_\odot}]$, $r_{_{0/s}} \ [\mathrm{kpc}]$, and $\rho_{_{0/s}} \ [\mathrm{gm \ cm^{-3}}]$.}
    \label{fig:IRCs}
  \end{center}
\end{figure*}

\begin{figure*}
 \begin{center}
  \includegraphics[angle=0,height=8.0truecm,width=8.0truecm]{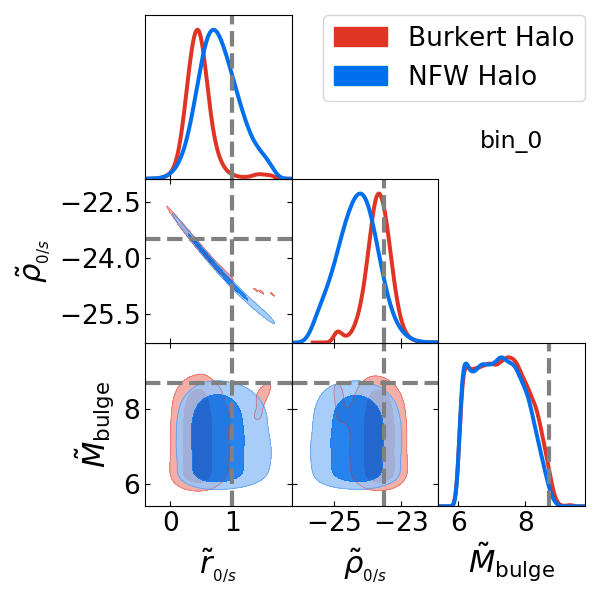} 
     \includegraphics[angle=0,height=8.0truecm,width=8.0truecm]{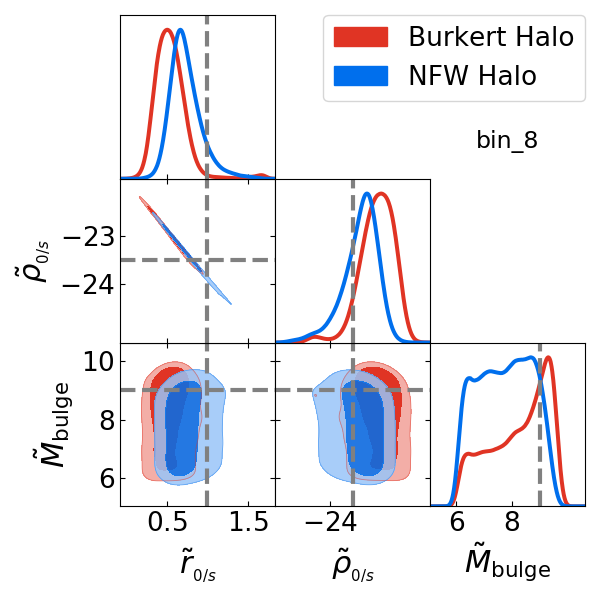}

  \includegraphics[angle=0,height=6.5truecm,width=8.5truecm]{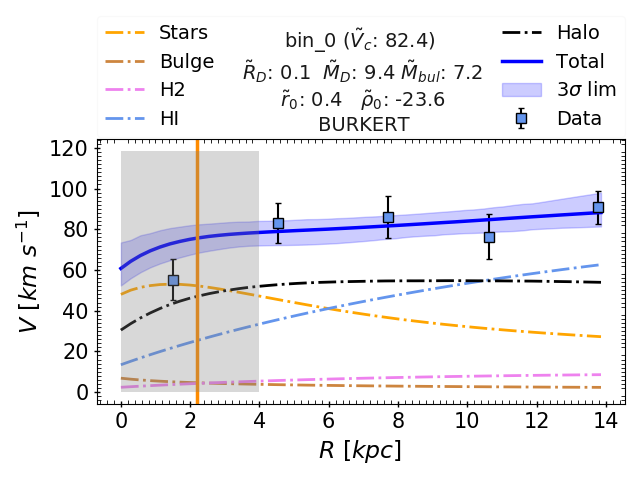} 
  \includegraphics[angle=0,height=6.5truecm,width=8.5truecm]{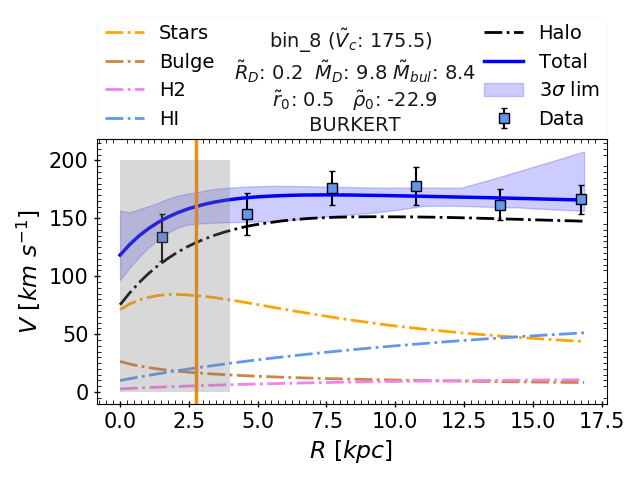}

  \includegraphics[angle=0,height=6.5truecm,width=8.5truecm]{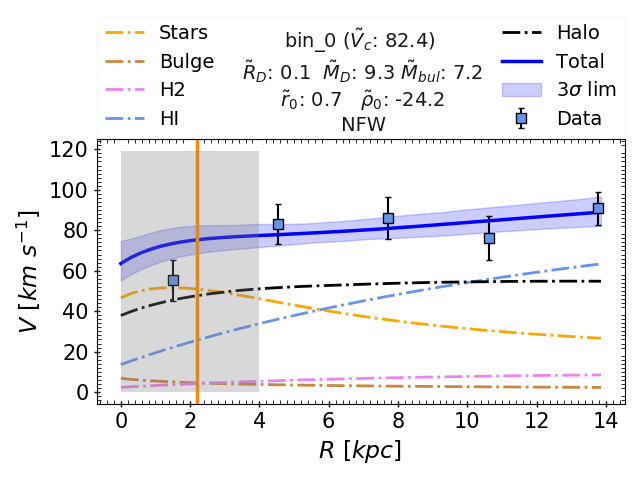}  
  \includegraphics[angle=0,height=6.5truecm,width=8.5truecm]{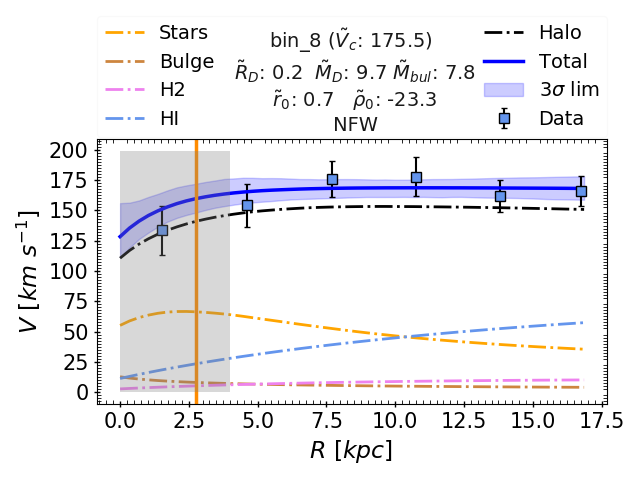}

    \caption{\underline{Coadded RC mass modeling}. {\em Row 1:} Posterior distributions (MCMC output) of the following estimated parameters: $\tilde{r}_{_{0}}$ and $\tilde{\rho}_{_{0}}$ for Burkert (red) and $\tilde{r}_{_{s}}$ and $\tilde{\rho}_{_{s}}$ for the NFW (blue) halo. The vertical and horizontal gray dashed lines show the initial guess for the parameters $\tilde{r}_{_{0/s}}$, $\tilde{\rho}_{_{0/s}}$, and $\tilde{M}_{bulge}$. The full posterior distribution of parameter space $\vec{\theta}$ of these example cases are shown in Figure~\ref{fig:SRCs_full}. {\em Rows 2 \& 3:} Best fit to the RCs and their velocity decomposition for the Burkert and NFW halos, respectively. In the RC panels, the blue square with error bars represents the observed data, the blue solid line shows the best fit to the data, and the blue shaded area represents the $3\sigma$ error in the fit. The decomposed velocity of stars, molecular gas, atomic gas, bulge, and dark matter is represented by yellow, violet, light blue, brown, and black dotted-dashed lines, respectively. The vertical orange line shows the effective radius and the gray shaded area represents the unresolved region of the RC. The best-fit value of parameter space $\vec{\theta}$ is given in the legend of the figures. All the parameters are printed in log, having physical units as follows: $\tilde{M}_{\mathrm{D}} \ [\mathrm{M_\odot}]$, $\tilde{R}_{\mathrm{D}} \ [\mathrm{kpc}]$, $\tilde{M}_{\mathrm{bulge}} \ [\mathrm{M_\odot}]$, $\tilde{r}_{_{0/s}} \ [\mathrm{kpc}]$, and $\tilde{\rho}_{_{0/s}} \ [\mathrm{gm \ cm^{-3}}]$.}
    \label{fig:SRCs}
  \end{center}
\end{figure*}

\begin{table*}
\centering
\begin{tabular}{|l|l|l|l|l|l|}
\hline
 & & \\
Parameter Range &  Case-1 (Individual RCs) & Case-2 (coadded RCs) \\
 (log scale, Units) & Prior: central value, $\sigma$ (dex) & Prior: central value, $\sigma$ (dex) \\
\hline
\hline
 & & \\
$8.0 \leq \log( M_{\mathrm{D}} \ \ [\mathrm{M_\odot}]) \leq 12.5$ & Gaussian Prior: $0.9 M_{\mathrm{star}}$, 0.25 dex  & Gaussian Prior: $0.9 \tilde{M}_{\mathrm{star}}$, 0.25 dex \\[1.2ex]

$-1.7 \leq \log(R_{\mathrm{D}} \ [\mathrm{kpc}]) \leq 1.7$ & Gaussian Prior: $R_D$, 0.25 dex  & Gaussian Prior: $\tilde{R}_D$, 0.25 dex \\[1.2ex]

$5.0 \leq \log(M_{\mathrm{bulge}} \ [\mathrm{M_\odot}]) \leq 11.5$ & Gaussian Prior: $0.1 M_{\mathrm{star}}$, 0.25 dex & Flat Prior \\[1.2ex]

$-2.0 \leq \log(r_{_{0/s}} \ [\mathrm{kpc}]) \leq 2.0$ & Flat Prior & Flat Prior \\[1.2ex]

$-26 < \log(\rho_{_{0/s}} \ [\mathrm{gm \ cm^{-3}}]) < -18$ & Flat Prior & Flat Prior \\[1.2ex]

\hline
\end{tabular} 
\caption{Mass modeling parameter range and their prior details. We note that for the Gaussian prior, central values were computed from the high resolution photometric data \citep[see][]{H17, GS21a}, and dispersion ($\sigma$) around the center is 0.25 dex for each photometric quantity. In the case of coadded RCs, the quantities are represented by a tilde on the top of them (e.g., $\tilde{R}_D$).}  
\label{tab:params}
\end{table*}

\subsection{Rotation curve modeling}
\label{sec:technique}
To perform the mass modeling of RCs, we defined a $\chi^2$ test statistic on our observed kinematics $V_o(R)$ and modeled the kinematics $V_m(R)$ as follows:
\begin{equation}
\label{eq:loglikehood}
 \mathcal{L}_{\rm{kin}}  \equiv  \prod_{k = 1}^{\rm{N}} \frac{1}{\sqrt{2\pi} \ \Delta V_o(R_{_k})} \exp \Big [ -\frac{1}{2} \Big( \frac{V_o(R_{_k}) - V_m (R_{_k})}{\Delta V_o(R_{_k})} \Big)^2  \Big].
\end{equation}
For each bin $k\leq \mathrm {N}$, we compared the theoretical prediction and the observed dataset by including the observational uncertainty $\Delta V_o(R_{_k})$ at each step. Our fitting procedure is based on Bayes' theorem:
\begin{equation}
\label{eq:bayes}
\mathcal{P}\left(\rm{Model}(\vec{\theta}) \ | \ \rm{data} \right) \propto \mathcal{L}_{\rm{kin}}\left( \rm{data} \ | \ \rm{Model}(\vec{\theta}) \, \right) \ \mathcal{P}_{0}\left(\rm{Model}(\vec{\theta}) \, \right),
\end{equation} 
where the posterior probability density function (p.d.f.) is sampled from the product of the prior probability distribution assigned to the set of model parameters $ \vec{\theta}$, with the  likelihood function reported in Equation~\ref{eq:loglikehood}, up to the overall normalization defining the so-called  evidence of the model. The general model of the RC under the Bayesian inference is defined by five parameters $\vec{\theta} = \{M_{\mathrm{D}}, R_{\mathrm{D}}, M_{bulge}, r_{_{0/s}}, \rho_{_{0/s}} \}$. Here, three parameters, namely $ M_{\mathrm{D}}, R_{\mathrm{D}}, $   and $ M_{bulge},$ constrain the baryons, and two parameters, namely $r_{_{0/s}} $  and $  \rho_{_{0/s}}$, describe the DM halo. The details of these parameters in various cases (studied in this work) of mass-modeling is given in Table~\ref{tab:params}. By employing Equation~\ref{eq:loglikehood} \& \ref{eq:bayes}, we performed the MCMC analysis exploiting the RC dataset discussed in Section~\ref{sec:data}. To perform the MCMC, we made use of the {\em emcee} package \citep{emcee}. For each object, we allowed 100 walkers to evolve for 10,000 steps, which gave us $10^6$ samples for analysis. We removed the first half of the Markov chains to account for the burn-in period and the remaining were used to draw the posteriors of parameter space $\vec{\theta}$. Lastly, we tested our mass modeling technique on synthetic RCs. Those are modeled as in Section~\ref{sec:models} using the realistic values for the parameter space $\vec{\theta}$. In Appendix~\ref{sec:test-MM}, we show that our MCMC approach provides an accurate fit to these synthetic RCs and that the retrieved parameter space $\vec{\theta}$ is consistent, within the $1\sigma$ uncertainty, with the input parameters.


\subsection{Fitting details}\label{sec:fitting-details}
As we briefly mentioned above, in the fitting procedure, we have five free parameters, given by $\vec{\theta} = \{M_{\mathrm{D}}, R_{\mathrm{D}}, M_{bulge}, r_{_{0/s}}, \rho_{_{0/s}} \}$, where three parameters, namely $ M_{\mathrm{D}}, R_{\mathrm{D}}$, and $M_{\mathrm{bulge}},$ constrain the baryons and two parameters, namely $r_{_{0/s}}$ and $\rho_{_{0/s}}$ describe the DM in the case of the Burkert and NFW halo. The exploration range of each parameter is motivated by the mass modeling of local galaxies \citep[see,][]{PS1996, Karukes2017, lapi2018, Chiara2019}, given in Table~\ref{tab:params}. As mentioned above, we performed RC modeling in two cases: {\bf case 1)} individual RCs, and {\bf case 2)} coadded RCs. In both cases, an initial guess on the parameters is the same and given as follows $\vec{\theta} = \{10, 0.3, 8.5, 1, -23.5\}$ (in dex). The prior on the DM parameters were kept flat in both cases. On the other hand, the baryonic parameters have different conditions for prior in different cases, each motivated by the need to treat the data robustly and without biases. In the case 1, we used the Gaussian prior in the log scale for $M_{\mathrm{D}}, R_{\mathrm{D}}$, and $M_{bulge}$ with a dispersion ($\sigma$) of 0.25 dex. In the log space, the unit is dex. Where the central values for $M_{\mathrm{D}}=0.9M_{\mathrm{star}}$, which at first approximation excludes the bulge. The values of $M_{\mathrm{star}}$ and $R_{\mathrm{D}}$ are taken from \citet{H17}. The bulge mass $M_{bulge}$ is one-tenth of the stellar mass, which is motivated from local SFGs \citep[for reference see][]{Bruce2014, Morselli2017}. In the case 2, we allowed the Gaussian prior only on $\tilde{R}_{\mathrm{D}}$, and $\tilde{M}_{\mathrm{D}}$, having a dispersion of 0.25 dex around the central value, and all other parameters kept a flat prior.  In this case, central values are the quantities binned in the coaddition process of RCs, as discussed  in Section~\ref{sec:CRCs}. Similar to the case 1, central value of $\tilde{M}_{\mathrm{D}}$ is $0.9 \tilde{M}_{\mathrm{star}}$.

We note that our modeling approach also requires other parameters in the fitting process, namely $M_{\mathrm{H2}}$, $M_{\mathrm{HI}}$, $R_{\mathrm{H2}}$, and $R_{\mathrm{HI}}$ (see Equation~\ref{eq:vtot}). In particular, $M_{\mathrm{H2}}$ and $M_{\mathrm{HI}}$ were calculated in each MCMC run for a given star formation rate and redshift ($M_{H2, HI }(SFR, M_{\mathrm{D}}, z)$) using the scaling relations given by \citet[][]{Tacconi2018} and \citet{Lagos2011}, respectively. The molecular gas scale length $R_{\mathrm{H2}}$ was determined from the $H\alpha$ surface brightness distribution, since ionized gas around a star is surrounded by molecular gas clouds. For a detailed calculation of $M_{\mathrm{H2}}$, $M_{\mathrm{HI}}$ and $R_{\mathrm{H2}}$, we refer the reader to \citet{GS21b}. Moreover, studies of local disk galaxies have shown that the surface brightness of the HI disk is much more extended than that of the H2 disk \citep[][see all Fig. 5]{Jian2010} see also \citep{Leroy2008, Cormier2016}. Therefore, we assume $R_{\mathrm{HI}}=2 \times R_{\mathrm{H2}}$, which is a rough estimate but still reasonable considering that at high-$z$ no information is available on the $M_{HI}$ (or $M_{H2}$) surface brightness distribution. Furthermore, it is important to note that these additional parameters, such as the star-formation rate (SFR), $M_{gas}$, and $R_{gas}$,  were fixed in the modeling procedure to the best of our knowledge.

While performing our mass modeling on individual RCs (i.e., Case-1), we visually (qualitatively) inspected the individual RCs and each of their individual fittings. We found that there are 28 perturbed RCs that we could not model, that is, we mass-modeled 228 individual RCs.\footnote{These 28 RCs have already been identified and reported in our previous work \citep[][see Appendix Fig. D1]{GS21b}. These kinds of perturbed RCs generally arise due to atmospheric turbulence while an observation or due to a disturbed morphology and the potential of a galaxy itself. However, we do not see these systems in merging state (inspected using HST imaging); therefore these are likely bad observations.} After the mass modeling of these RCs, we visually inspected each MCMC output (posteriors) and found that the 92 objects have hit the prior boundaries, either for DM parameters or for baryons (in particular $M_{\mathrm{D}}$). This tells us that the fitting of these 92 objects is prior-driven, so we discarded them from the individual analysis. Thus, we ended up analysing only 136 individual RCs. It is important to note, however, that in case 2, we used up all of the 228 RCs, making 16 coadded RCs (see Section~\ref{sec:CRCs}), which were mass modeled using the aforementioned prescription.

\section{Results \& analysis}
\label{sec:results}
As described in Section~\ref{sec:models} \& \ref{sec:technique}, by inputting the individual RCs (case 1) and the coadded RCs (case 2), we constrained the parameter space $\vec{\theta}$ and disentangled the various  components of each RC. With this information, we estimated the quantities $R_{\mathrm{vir}}$, $M_{\mathrm{vir}}$, $c_{vir}$, and $\fdm$. In the following section, we present these results.

                      
\subsection{Mass modeled RCs}
\label{sec:MM-RCs}
Here, we present the results of mass modeling in case 1 and case 2. Some examples of their MCMC outputs (posteriors) and best fit to the RCs are shown in Figure~\ref{fig:IRCs} \& \ref{fig:SRCs}, respectively. We note that in both cases, the Burkert and NFW halo models provide similarly good fits (in terms of minimum $\chi_{red}^2$) to the data. The rationale is that both halo models predict a similar DM density profile in the outskirts of the galaxies and differ only in the very inner region ($<2$ kpc), where our data are unresolved. Therefore, we cannot exclude one model over the other just by looking at the rotation curve fitting. 

\begin{figure*}
  \begin{center}
    \includegraphics[width=\columnwidth]{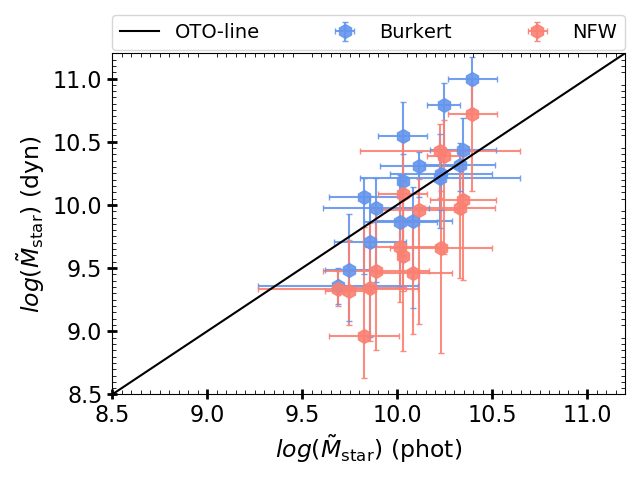}
    \includegraphics[width=\columnwidth]{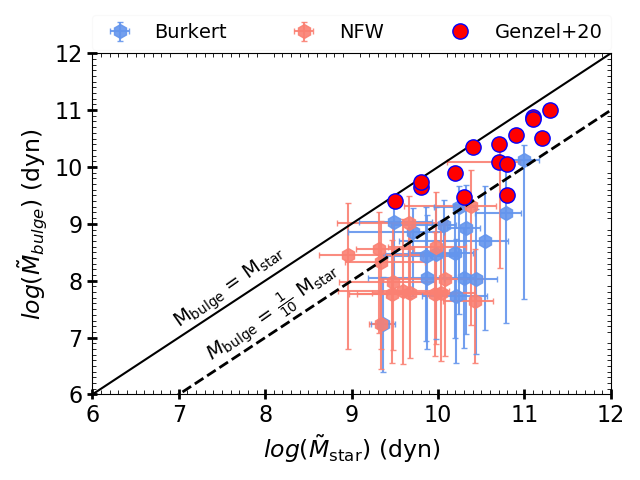}
    
    \caption{Dynamically derived integrated quantities. {\em Left panel:} Comparison of the photometric and dynamical stellar masses in the Burkert (blue) and NFW (pink) halo profiles for coadded RCs. The black solid line shows the one-to-one relations. We note that the dynamical stellar masses obtained assuming the Burkert halo agree very well with the photometric stellar masses. In contrast, the dynamical stellar masses obtained with the NFW halo are lower than the photometric stellar masses by a factor of 0.5 dex. {\em Right panel:} Dynamically derived bulge masses versus stellar masses, in the case of Burkert (blue) and NFW (pink) halo profiles for coadded RCs. We compared our results of the same redshift range ($z=0.6-1.2$) and a similar stellar mass sample with a pioneering study by \citet[][: red filled circles]{Genzel2020}, and we found that in our sample $B/T$ is never higher than $\sim 0.3$, which is different from what is reported by \citet{Genzel2020}.}
    \label{fig:Mdyn-Mphot-Mbulge}
  \end{center}
\end{figure*}

Next, we focused on comparing the physical properties of galaxies, such as the total stellar and bulge masses. In particular, we compared the physical properties derived from case 2 because this case gives us a smooth distribution of RCs that reduce the effect of minor observational errors (perturbations caused by bad data points), which make the mass modeling more impeccable than in case 1. In the left panel of Figure~\ref{fig:Mdyn-Mphot-Mbulge}, we compare the dynamically derived stellar masses from the Burkert and NFW halo of case 2 with the photometric stellar masses. As we can see, the dynamical stellar masses from the Burkert halo agree with the photometric stellar masses with an intrinsic scatter of 0.3 dex. In contrast, the dynamical stellar masses derived from the NFW halo are a factor of 0.5 dex lower than the photometric stellar masses, with an intrinsic scatter of 0.43 dex. This suggests that the NFW halo most likely suppresses the baryon content, which could be problematic while deriving the DM fraction.

\begin{figure*}
  \begin{center}
   \includegraphics[angle=0,height=6.5truecm,width=9.0truecm]{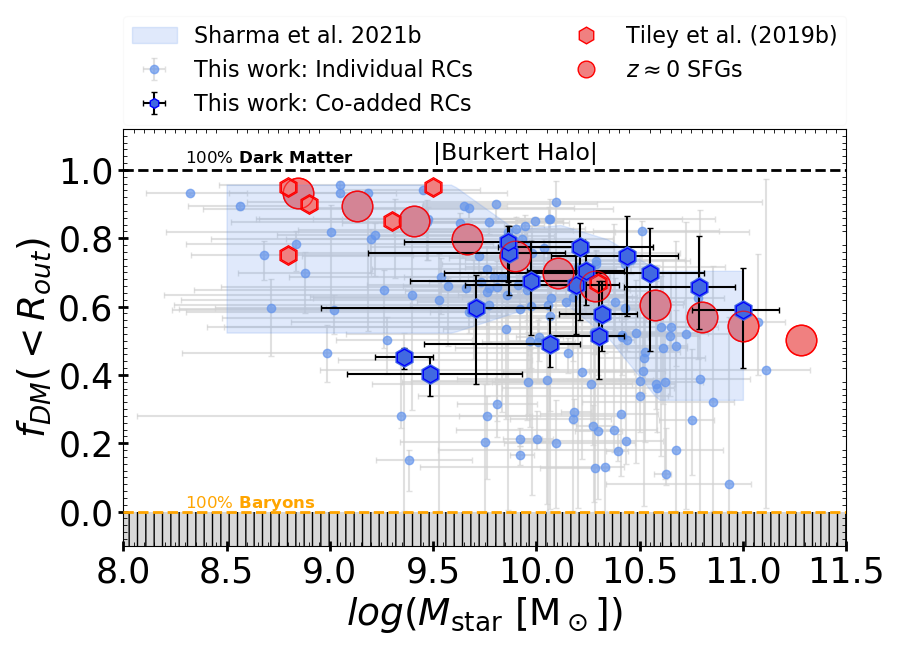}       
  \includegraphics[angle=0,height=6.5truecm,width=9.0truecm]{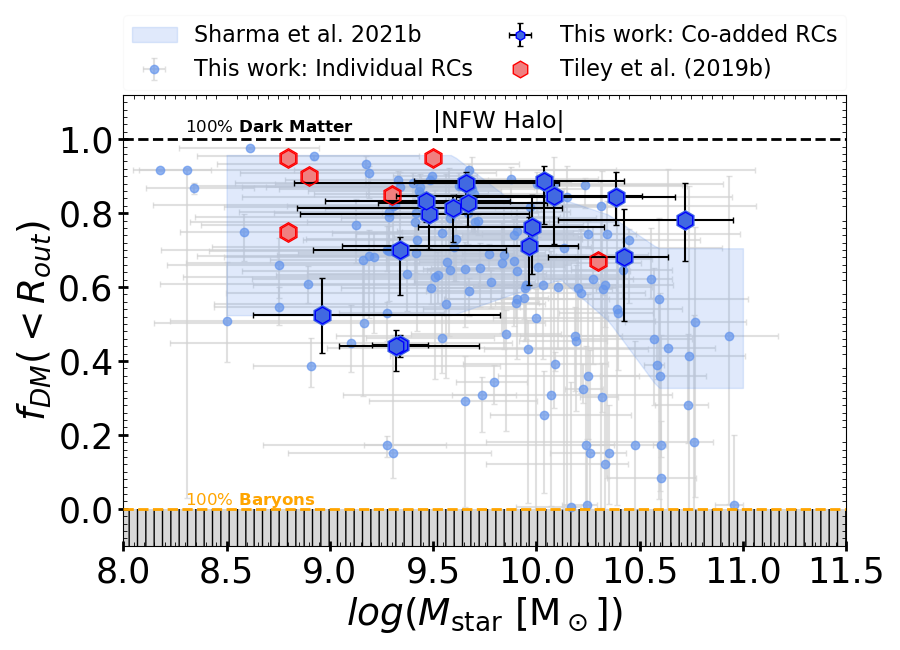} 
    
    \caption{DM fraction within $R_{\mathrm{out}}$ for the Burkert and NFW halos, on the left and right, respectively. The color coding in both panels are the same for the individual RCs (blue filled circles), coadded RCs (blue hexagons), local SFGs (coral red circles), and high-$z$ SFGs (coral red hexagons). The size of the makers in local SFGs includes their errors. The blue shaded area represents the model-independent DM fraction study that we previously conducted in \citet[][]{GS21b}. 
    }
    \label{fig:fdm-out-obs}
  \end{center}
\end{figure*}

Moreover, in both cases (with both halo models), we found it difficult to constrain the bulge mass, and often ended up with a broad distribution. However, the bulge is never heavier than $0.3 \times M_{\mathrm{star}}$ (see right panel of Figure~\ref{fig:Mdyn-Mphot-Mbulge}) and, therefore, it does not significantly affect the inferred DM fraction (see Figure~\ref{fig:fdm-out-obs}). In fact, the DM has emerged as a dominant quantity, especially in the outskirts\footnote{Here, outskirts refer to the region of galaxies beyond its effective radius.}, which is very similar to the local SFGs \citep{PS1996, GS21b}. We also found that for low mass objects ($M_{\mathrm{D}}< 10^{9.5} \ \Ms$), the atomic gas kinematics (HI) dominate the baryonic component; whereas for massive objects, the stellar kinematics prevail.


\subsection{DM fraction}
\label{sec:fdm}
In Figure~\ref{fig:fdm-out-obs} we show the results of the DM fraction within Rout ($=2.95 R_e$) inferred from our mass modeled RCs, in case 1 and case 2, for the Burkert and NFW halo. Firstly, we noticed that both halo models in both cases have shown that the majority ($\sim 70\%$) of galaxies are DM-dominated in the outskirts. Secondly, our current results are in very good agreement with our previous study \citep{GS21b}, where we determined the DM fraction using a halo model-independent approach. This confirms that the majority of SFGs at $z\sim 1$ have DM-dominated outer disks (up to $3\times R_e$).

In the case of individual RCs, in both halo models, nearly 20\% of objects appear to be deficient in DM (i.e., $f_{\mathrm{DM}}(<R_{out})<50\%$). We also note that the DM fraction derived from the individual RC modeling has a higher uncertainty in both halo models, while it is significantly reduced in the case of the coadded RCs. This could be due to the low resolution in the data (nonsmooth $V(R)$), which is penalized by the large uncertainty. On the other hand, coadding RCs statistically reduces the effect of the random noise introduced by the bad data points and provides us with a smooth RC, which can be easily dynamically modeled, and hence yields a relatively small uncertainty on the measured parameters. 

As we discussed in Section~\ref{sec:MM-RCs}, dynamical modeling assuming a NFW halo profile tends to suppress stellar mass (see left panel of Figure~\ref{fig:Mdyn-Mphot-Mbulge}). Therefore, the DM fraction obtained in this case is higher in comparison to the Burkert halo case (see right panel of Figure~\ref{fig:fdm-out-obs}). The latter argument is clearly visible from a coadded RC (blue hexagons) analysis. This suggests that dynamical modeling with the NFW halo requires precise baryon information, while the Burkert halo does not and it is still able to constrain the physical properties of galaxies, for example stellar masses. This is the reason that the $\fdm$-$M_*$ relation obtained with the Burkert halo matches that inferred in \citet{GS21b} very well for which a halo model-independent approach was used (blue shaded region in Figure~\ref{fig:fdm-out-obs}). However, this cannot be a strong piece of evidence to discard one model over the other because measurements from NFW models are still in good agreement within the 1$\sigma$ uncertainty.  

              

\begin{figure}
  \begin{center}
    \includegraphics[width=\columnwidth]{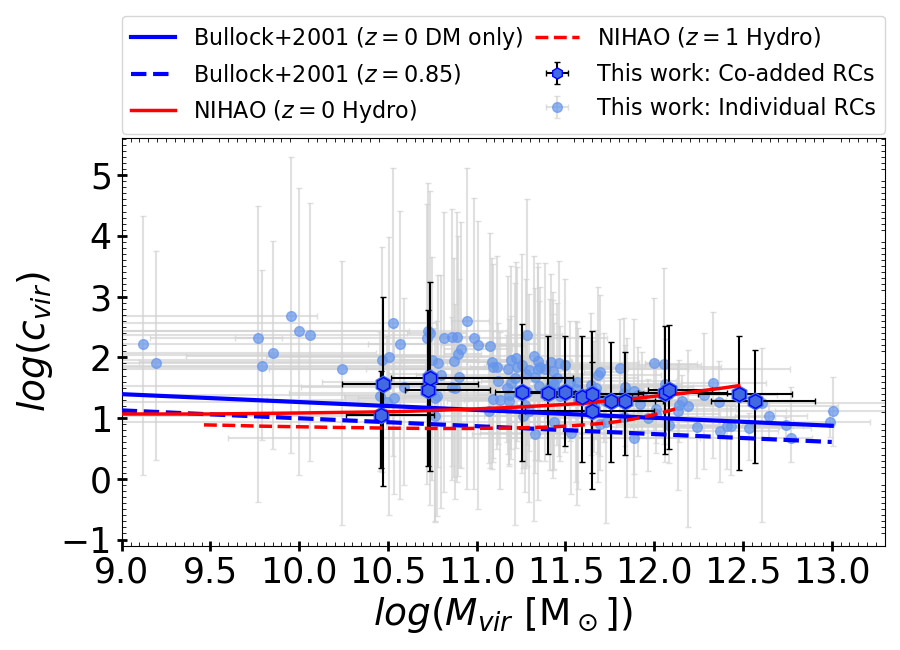}
    \caption{Concentration vs. virial mass relation for individual and coadded RCs (or galaxies) in blue circles and hexagons, respectively. The blue and red, solid and dashed lines represent the $c-M_{\mathrm{vir}}$ relation of DM-only simulations \citep{Bullock2000} and the NIHAO suite of hydrodynamical cosmological simulations \citep{Wang2015, JF2020}, for $z= 0$ and $z\sim 1$, respectively.}
    \label{fig:c-mvir}
  \end{center}
\end{figure}

\begin{figure}
  \begin{center}
    \includegraphics[width=\columnwidth]{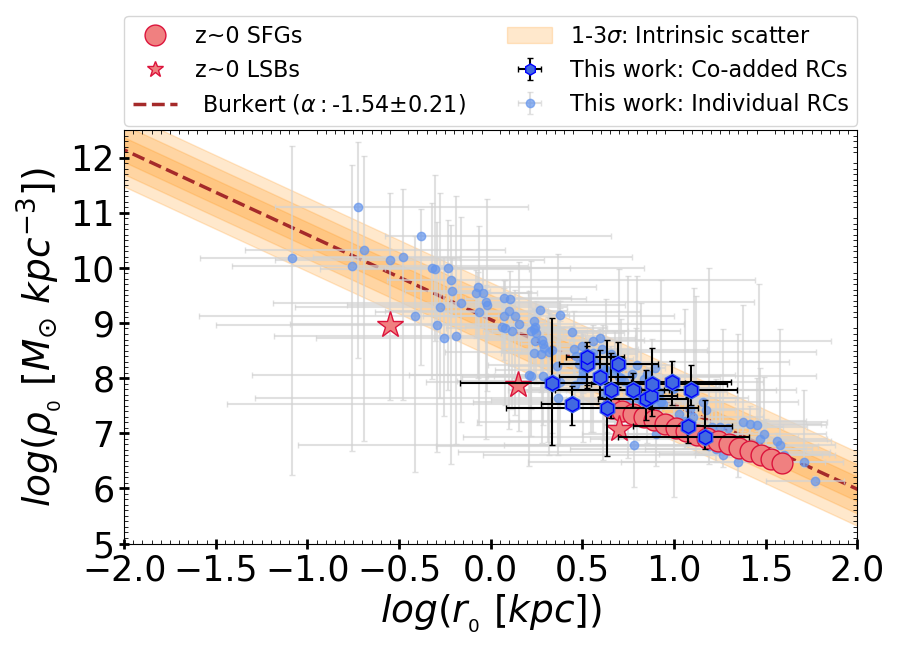}
    \caption{DM core density vs. core radius relation of local and high-$z$ galaxies. The blue circles and hexagons represent the $z\sim 1$ individual and coadded RCs, respectively. The coral red circles and stars represent the local star-forming and low surface brightness galaxies \citep{PS1996, Chiara2019}, respectively. The brown dashed line is the best fit $\rho_{_0} \propto r_{_0}^{\alpha}$ to $z\sim 1$ data with a slope $\alpha=-1.54$, and the orange shaded area represents the $1-3\sigma$ intrinsic scatter in the relation.}
    \label{fig:r0-rho0}
  \end{center}
\end{figure}

\begin{figure*}
  \begin{center}
   \includegraphics[angle=0,height=15.0truecm,width=18.5truecm]{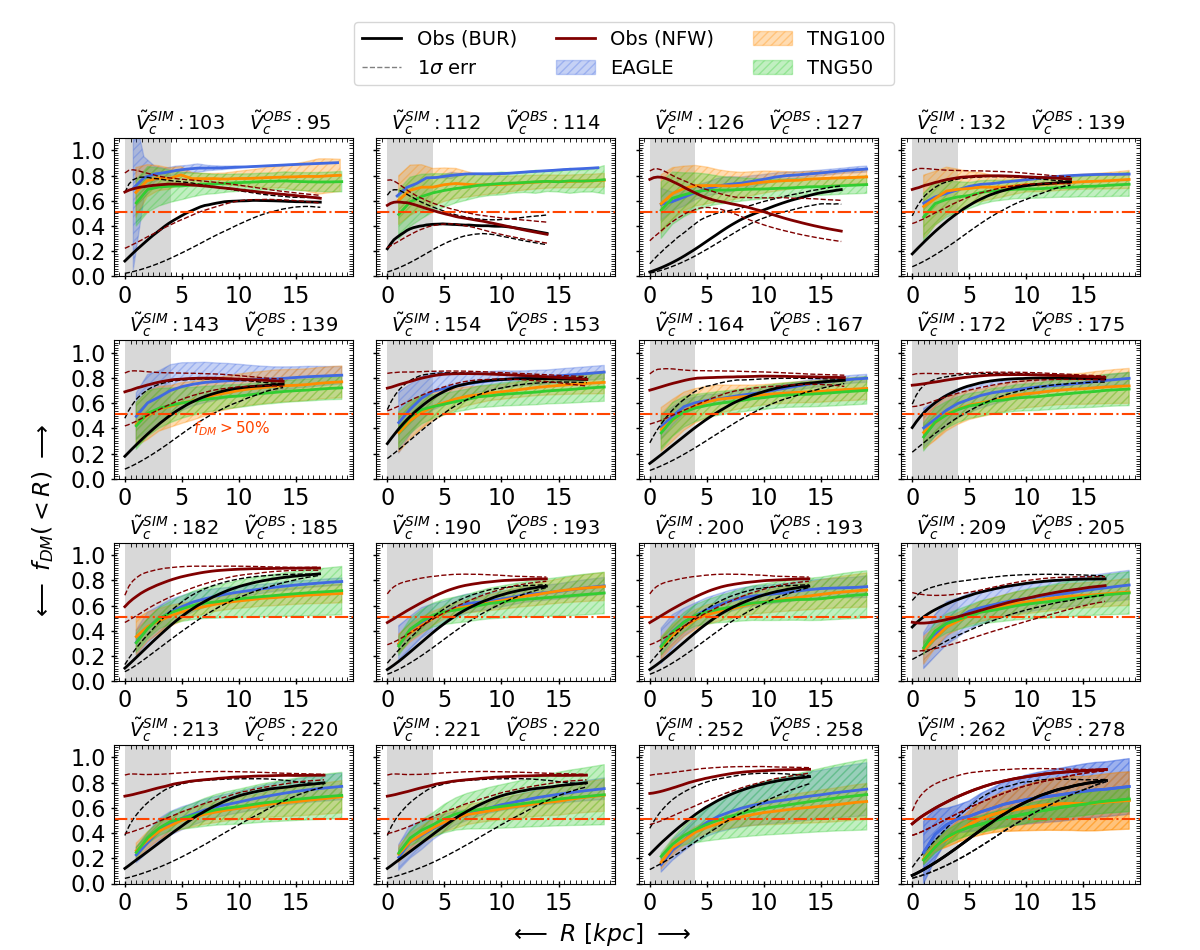}       
    
    \caption{DM fraction as a function of the radius in various velocity bins of observed and simulated galaxies. The central value of $f_{\mathrm{\tiny{DM}}}(< R)$ is shown by solid lines in each case. The color code is the same in all panels and is as follows: EAGLE (blue), TNG100 (orange), TNG50 (green), observations in the case of the Burkert halo (black), and NFW halo (brown). In the simulations, the $1\sigma$ error in each measurement is represented by a shaded area; whereas in the observations, it is a dashed line following the same color code. The gray shaded area in each panel represents the unresolved region in the observations. The horizontal red dotted-dashed line in each panel shows the 51\% DM fraction. The central value of the circular velocity of each bin is printed at the top of each panel, where $\tilde{V}_c^{ SIM }$ and $\tilde{V}_c^{ OBS }$ specify the simulations and observations, respectively.}
    \label{fig:sims}
  \end{center}
\end{figure*}

\begin{figure*}
  \begin{center}
	\includegraphics[angle=0,height=10.7truecm,width=15.0truecm]{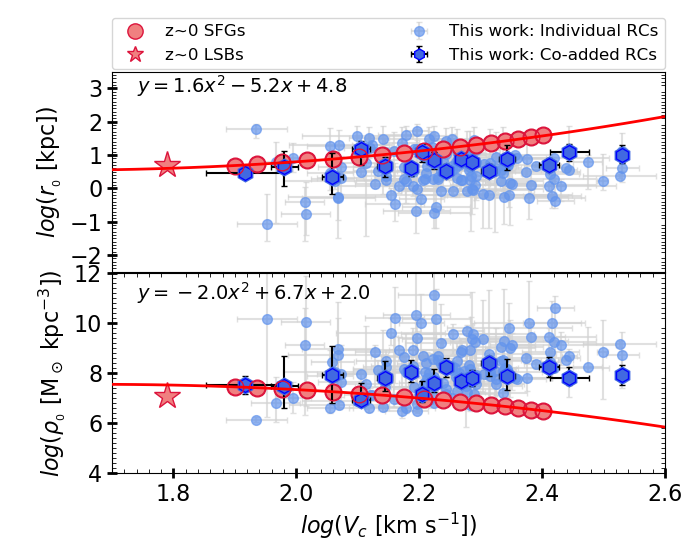}
    \includegraphics[angle=0,height=10.7truecm,width=15.0truecm]{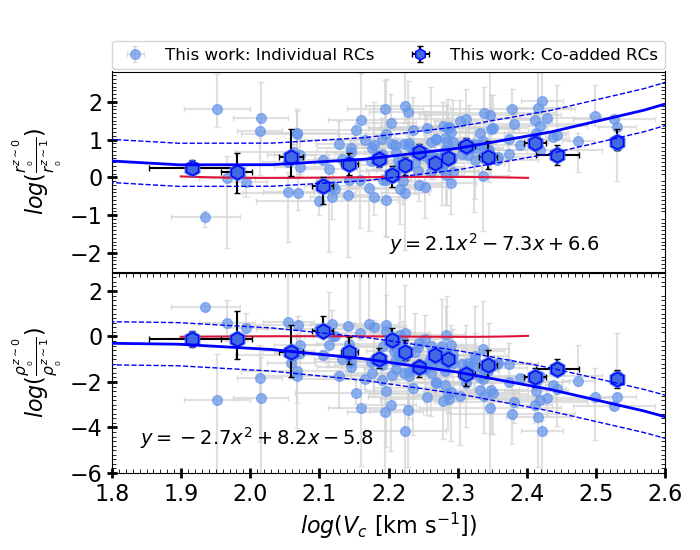}
    
    \caption{Structural properties of the DM halo at $z\sim 1$. {\em Top:} DM core radius and core density as a function of the circular velocity shown in the upper and lower panels, respectively. The individual and coadded RC measurements are represented by small blue circles and hexagons, respectively. The comparison sample of local LSBs \citep{Chiara2019} and SFGs \citep{PS1996, lapi2018} are represented by the coral red stars and circles, respectively. The best-fit line (second-order polynomial) of the local SFGs and LSBs is represented by a red line, and the equation is printed in the upper left corner of each panel. As we can notice, the local SFGs lie above in the $r_{_0} - V_c$ plane and below in the $\rho_{_0}-V_c$ plane with respect to the $z\sim 1$ SFGs. This is due to the fact that DM cores of $z\sim 1$ SFGs are smaller and denser than their local counterparts, indicating an evolution in the structure of DM halos over cosmic time. {\em Bottom:} Ratio of DM core radius and density at $z\approx 0$ to $z\sim 1$ as a function of circular velocity. This plot shows the expansion factor of DM cores at $z\sim 1$ with respect to local SFGs. The color codes for data points are similar to those in the top panel. The blue solid line shows the best fit to the data, obtained by fitting the second order polynomial function printed in both panels. The dashed blue lines followed by a solid line shows the 1$\sigma$ intrinsic scatter in the relations.}
    \label{fig:halo-props}
  \end{center}
\end{figure*}


\subsection{Structural properties of DM}
\label{sec:SP-DM}
With the information on the DM halo parameters $r_{_{0/s}}$ and$ \ \rho_{_{0/s}}$, we derived the physical properties of the galaxies, such as $R_{\mathrm{vir}}$ and $ M_{\mathrm{vir}}$, and the concentration in the case of the NFW halo. Using the latter quantities, we compared the $c_{vir}-M_{\mathrm{vir}}$ relation of our sample with DM-only simulation \citep{Bullock2000} and the NIHAO hydrodynamical simulation \citep{Wang2015}, as shown in Figure~\ref{fig:c-mvir}. We found that the concentration in $z\sim 1$ SFGs goes in the opposite direction, which challenges the simulations. In other words, both the DM-only and hydro simulations predict the halo concentration which decreases with increasing redshift (see red and blue dashed lines in Figure~\ref{fig:c-mvir}). The observations show a relatively high concentration at $z\sim 1$, which is somewhat in tension with current simulations, although they remain consistent with them within the $1\sigma$ uncertainty. We remark that the concentration was derived as $c_{vir}=R_{\mathrm{vir}}/r_{s}$, where the determination of $R_{\mathrm{vir}}$ requires a huge extrapolation. Therefore, the concentration in this work is not as accurate as it can be derived from HI RCs, where data are available over more than a 50 kpc radius. Thus, although the NFW halo gives conflicting results with respect to the simulations, we cannot falsify its presence in nature.

On the other hand, the relation between the core density and core radius, associated with the Burkert profile, emerge very similar to the local SFGs, with a similar slope, as shown in Figure~\ref{fig:r0-rho0}. However, we found that high-$z$ galaxies have denser DM cores, as they lie above the local SFGs and low surface brightness galaxies (LSBs). At the same time, the size of their DM core is smaller with respect to local SFGs, but equivalent to the core of LSBs. This suggests that the structural properties (i.e., structure) of DM have evolved over cosmic time. In particular, DM cores at $z\sim 1$ are smaller by a factor of 0.3 dex and denser by 1.5 dex.


\section{Comparison with simulations}
\label{sec:sims-comp}
In this section, we compare the mass modeling results with the current state-of-the-art galaxy simulations: EAGLE, TNG100, and TNG50. The details of their sample selection and RC data can be found in Appendix~\ref{sec:sims-data}. In particular, we compare our observed coadded RCs with the simulated coadded circular velocity curves. Briefly, in simulations we know the exact total circular velocity of the galaxy as a function of radius ($V_{c}(R)$), and its various components, such as $V_{\mathrm{star}}(R)$,  $V_{\mathrm{\tiny {DM}}}(R)$, and $V_{\mathrm{gas}}(R)$. From observations we only know the $V_{c}(R)$, which we mass modeled to disentangle the various components (see Section~\ref{sec:models}). That is to say, in both cases, observations and simulations, we know the $V_{c}(R)$ and $V_{\mathrm{DM}}(R)$. This allowed us to compute the DM fraction of galaxies as a function of radius using Equation~\ref{eq:fdm}. In Figure~\ref{fig:sims} we show the results of $\fdm(< R)$ in the aforementioned simulations. In particular, we show 16 velocity bins, in which the first four bins have a few galaxies (in particular EAGLE simulations), and the remaining 12 bins have enough simulated data (in all simulations) to bin and obtain the coadded RCs; for details, readers can refer to Table~\ref{tab:sims-params}. It is noticeable that all simulations show DM-dominated galaxies from 5 kpc outward. In other words, the simulated galaxies have 20\% (minimum) to 50\% DM within the effective radii, and they are DM-dominated in the outskirts. These results are very similar to the observed SFGs at $z\sim 1$, reported previously in \citet{GS21b}. We remind the reader that galaxies within effective radii are known to have a lower DM fraction because baryon dominates the inner region of galaxies \citep[][and references therein]{PS1996}. If one has to look for DM, it should be found in the outskirts ($> 2 \times R_e$), as clearly shown in Figure~\ref{fig:sims}.

To compare the simulations with the observations, we superimposed the $\fdm (< R)$ of the observed coadded RCs in Figure~\ref{fig:sims}, in the case of the Burkert (black) and NFW (brown) halos. Since the observations in the inner region are not resolved, we mark the inner region (0-4 kpc) as a gray shaded area and avoid drawing any conclusions there. It is important to notice that we compare the observed and simulated coadded RCs only when $V_c^{ SIM } \approx V_c^{ OBS }$, as shown at the top of each panel. First, we noticed that in observations, we have roughly three coadded RCs, corresponding to $log(M_{\mathrm{star}} \ [M_\odot]) \approx 9.2, 9.4, 10.0$, with a low DM fraction (i.e., $\fdm$ < 50\%) in the outskirts (for Burkert: $V_c^{Obs} \sim 83, 114 \ \mathrm{km/s} $ and NFW $V_c^{Obs} \sim 83, 114, 127 \ \mathrm{km/s}$). We found that these RCs have a gas-dominated outer disk, which we could not find in the simulations. This could be either due to limitations of simulations, or scaling relations that we used in gas mass estimates. Despite very low velocity bins ($V_c <130$ km/s) for both Burkert and NFW halos, we noticed DM-dominated outer disks. Second, the NFW and Burkert halo results at $R>10$ kpc are very similar in intermediate velocity bins ($V_c^{ SIM } \approx 130-170 \ \mathrm{km \ s^{-1}}$) and they agree well with the simulations within a $1\sigma$ uncertainty. However, when we go toward the higher velocity range, the NFW halo starts to show a deviation from the simulations. In particular, the Burkert halo and the simulations have almost the same DM fraction between the 10-15 kpc radius, which is different from the NFW halo. In short, the Burkert profile seems to represent the simulations more accurately. 

To observe the largest differences, we emphasize that a reader should look at the DM fraction between the 5-10 kpc radius in Figure~\ref{fig:sims}, where we can clearly notice that the NFW halo allows for more DM content than for the Burkert halo, which we do not even see in the simulations. In other words, the DM fraction derived from the Burkert halo model is close to the one seen in hydrodynamical simulations for the majority of circular velocity bins, while NFW yields a relatively high DM fraction. The latter could be a consequence of suppressing the stellar mass which we have already noticed from the mass modeling of observations in Section~\ref{sec:MM-RCs} \& \ref{sec:fdm},  and Figure~\ref{fig:Mdyn-Mphot-Mbulge} \&  \ref{fig:fdm-out-obs}. 



\section{Discussion}
\label{sec:discussion}
To fit the individual RCs through a mass modeling approach, one needs to use the informative prior on the baryons ($M_{\mathrm{D}}, M_{bulge}$, and $R_{\mathrm{D}}$), otherwise the parameter space $\vec{\theta}$ is highly degenerated. In the case of coadded RCs, on the other hand, we could achieve a good fit only by specifying two baryonic parameters, which are the stellar disk scale length and stellar mass in this work. This case also allowed us to adopt the flat prior on bulge mass, which is an unknown quantity (and unresolved), and thus we can determine it dynamically (see Section~\ref{sec:MM-RCs} and Table~\ref{tab:params}). Based on RC fitting and structural properties of DM, we cannot robustly validate or invalidate the NFW halo model, but tension exist. For instance, a galaxy cannot have a stellar mass lower than what is measured from photometry, and if so, it misleads the mass budget of a galaxy. The latter happens in the case of the NFW halo model fitting, where mass modeling yields stellar masses that are lower than that inferred from the observed photometry. This results in a high DM fraction that can be seen in the  right panel of Figure~\ref{fig:fdm-out-obs}, which we do not see when using the Burkert halo, nor in the simulations (EAGLE, TNG100, and TNG50; see Figure~\ref{fig:sims}). Providing the various results and comparisons of the NFW and Burkert halo in the case of individual and coadded RCs, we found that our current sample favors the Burkert halo model and also agrees quite well with the simulations. Moreover, the recent studies by \citet{Genzel2020} and \citet{Bouche2021} also found a cored profile in majority of their SFGs at $z\sim 1$. Therefore, we focus on discussing the Burkert halo results in what follows.

We begin by referring to our earlier study of the DM fraction \citep{GS21b}, in which we used a halo model-independent approach to determine the $\fdm$ from RCs, and showed that the galaxies are DM-dominated until the last point of observation ($1-3 \ R_e$). We found that our earlier finding agrees very well with our current results of the DM fraction obtained via mass modeling (see the left panel of Figure~\ref{fig:fdm-out-obs}). These results are  also consistent with the previously studied local and high-$z$ SFGs, indicated by coral red circles and hexagons, respectively (see  Figure~\ref{fig:fdm-out-obs}). In this work, we do not compare the DM fraction within $R_e$ because it is beyond the scope of this paper. However, it is done in our previous work \citep[][see Fig. 5]{GS21b}, where we have compared $\fdm (<R_e)$ of our sample with the findings of \citet{Genzel2020}. We have shown that $\sim 25\%$ of their $z\sim 1$ sample agrees well with our work, and the rest of the sample shows a low DM fraction (i.e., baryon-dominated disk) as their sample is biased toward massive galaxies. Now, with dynamical mass-modeling we refined our previous work, which allowed us to compute the DM fraction as a function of the radius of a galaxy (see Figure~\ref{fig:sims}). As one can notice, the majority of the sample shows DM-dominated outer disks (from 5-15 kpc, shown by a horizontal orange dotted-dashed line), which falsifies the general remark of a low and modest DM fraction in $z\sim 1$ SFGs, made in \citet{Genzel2020}. Meanwhile, our new results of $f_{\tiny{DM}} (<R)$ match with the simulations (see Figure~\ref{fig:sims}). Furthermore, we would like to mention that the lower velocity end ($V_c < 130$ km/s) of our pilot study is less understood, firstly because we have very few (one to three) galaxies in simulations, which do not allow us to perform an accurate statistical analysis and, secondly, scaling relations that allow us to estimate the gas content from observed galaxies are likely not applicable on small systems (i.e., $V_c < 130$ km/s). Therefore, in the future, we will improve upon these two aspects using ALMA observations, upcoming HI-surveys, and by enlarging the simulated sample.

In one of our previous works \citep{GS21a}, we have shown that the progenitors of present-day SFGs at $z\sim 1$ have similar rotation curves. Given the similarity in the RCs of $z\sim 1$ and $z= 0$, and acknowledging the fact that the circular velocity is an indicator of the total mass of the system, we have concluded that the total mass within the optical extent of galaxies does not evolve from $z\sim 1$ to $z= 0$, while the stellar mass distribution (traced by the light profile) evolves. To bring together our previous and current findings, we compared the structural properties of the DM halo at $z\sim 1$ and $z\approx 0$ in the circular velocity plane (Figure~\ref{fig:halo-props}).
We would like to remark that the $r_{_0}-V_c$ and $\rho_{_0}-V_c$ relations of local galaxies are well fitted by second order polynomials (see upper panel Figure~\ref{fig:halo-props}), which are given by the following:
\begin{equation}
\label{eq:rc-Vc-z0}
\log r^{z\sim 0}_{_0} = 1.63 \ \ (log V_c)^{2} - 5.24 \ \log V_c + 4.76,
\end{equation}

\begin{equation}
\label{eq:rho-Vc-z0}
\log \rho^{z\sim 0}_{_0} =  -1.99 \ \ (log V_c)^{2} +6.67 \ \log V_c +2.04.
\end{equation}

Therefore, to visualize the differences in the structural properties of DM at different epochs, we used Equation~\ref{eq:rc-Vc-z0} and \ref{eq:rho-Vc-z0} as a benchmark and plotted the ratios: $\frac{r_{_0}^{z\sim 0}}{r_{_0}^{z\sim 1}}$ and $\frac{\rho_{_0}^{z\sim 0}}{\rho_{_0}^{z\sim 1}}$, as a function of circular velocity in the lower panel of Figure~\ref{fig:halo-props}. This immediately shows us how the structural properties of DM differ at $z\sim 1$ with respect to $z\approx 0$. First, we find that the DM cores at $z\sim 1$ are smaller than their local counterparts by a median factor of 0.3 dex. Second, the density of DM cores at $z\sim 1$ is 1.5 dex higher than at $z\approx 0$. It is also very clear that these two ratios correlate very well with the circular velocity of the galaxies. We refer to these ratios as the expansion factors of the core radius and density, which are both a function of the circular velocity and well fitted by the following:
\begin{equation}
\label{eq:exp-rc}
\log \Big( \frac{r_{_0}^{z\sim 0}}{r_{_0}^{z\sim 1}} \Big) = 2.1 \ \ (log V_c)^2 - 7.3 \ \log V_c + 6.6 \ \ \ \ \ \lfloor{\pm 0.6 \ \mathrm{dex}} \rceil ,
\end{equation}

\begin{equation}
\label{eq:exp-rhoc}
\log \Big( \frac{\rho_{_0}^{z\sim 0}}{\rho_{_0}^{z\sim 1}} \Big) = -2.7 \ \ (log V_c)^2 + 8.2 \ \log V_c -5.8 \ \ \ \ \ \lfloor{\pm 0.9 \ \mathrm{dex}} \rceil .
\end{equation}
The Equation~\ref{eq:exp-rc} \& \ref{eq:exp-rhoc} are accurate in relating the structural properties of DM to baryonic matter. Plotting these relations in the lower panel of Figure~\ref{fig:halo-props} clearly shows us that at $z\sim 1$, objects that have a higher circular velocity are denser in nature and more compact in size (in terms of DM cores), relative to $z\approx 0$ SFGs. In other words, objects with a higher circular velocity expand their DM cores more over cosmic time, compared to objects with a lower circular velocity. This means that the circular velocity plays a crucial role in the galaxy evolution. We report that the intrinsic scatter in relation~\ref{eq:exp-rc} \& \ref{eq:exp-rhoc} is about 0.6 and 0.9 dex, respectively. In addition to this, we also found that the galaxies in our sample have a typical DM core size of 3-10 kpc, which is a factor of 2.5 smaller than those reported in \citet{Genzel2020}. This discrepancy is likely due to the different methods and corrections applied in deriving the intrinsic RCs, which certainly affect the determination of the halo parameters. For example, most of the \citet{Genzel2020} RCs are declining, that is they are baryon-dominated systems. Therefore, the mass-modeling of these RCs yields a larger core radius of the DM halo and consequently a low concentration parameter.

The discussion above suggests that DM is most likely responding to the baryonic processes taking place in the inner region of galaxies throughout cosmic time. This also implies that if theory is correct and that early DM halos are cuspy (for instance at $z> 2$), they must have changed over cosmic time under the influence of galactic processes (e.g., AGN and supernova feedback as well as dynamical friction), which has been previously suggested by many authors such as \citet{Navarro1996, Pontzen2012, Nipoti2015, read2016, Read2018}, and \citet{Lazar2020}. Although, the difference in the size and density of the DM core between $z\sim 1$ and $z\approx 0$ is small, it is nevertheless remarkable and could have a very strong impact on the identification of the nature of DM and building the models for galaxy evolution. In short, our results suggest that the structure of DM halos expand over cosmic time. Based on these results, one can easily interpret that if the DM cores are denser at $z\sim 1$, then the virialization scale ($R_{vir}$) of these systems should also be smaller with respect to the locals. The latter argument can be further tested using the forthcoming observations of neutral hydrogen in the outskirts of galaxies \citep{Baker2018}. 

Furthermore, the results obtained with the Burkert halo model are consistent with the simulations. Therefore, so far we suggest that the rotation-dominated star-forming galaxies in the simulations and observations are very likely to have a DM core at the center of the halo. However, the inference of DM cores is very indirect in this study due to limited resolution in the inner region of galaxies. To confirm our findings, we still require high resolution and good S/N data in the inner region of the galaxies. That can possibly be obtained with the upcoming James Webb Space Telescope (JWST) and in future with the Extremely Large Telescope (ELT).

\section{Summary \& conclusions}
\label{sec:summary}
In this work, we have studied the structural properties of the DM in $z\sim 1$ star-forming disk-like galaxies. We first mass-modeled the individual and coadded rotation curves, assuming that the baryons are distributed in the \citet{Freeman1970} disk and that DM is modeled with the Burkert and NFW halo profiles. Then the mass modeling results in the Burkert and NFW cases were critically examined, interpreted, and compared with current state-of-the-art galaxy simulations. We found that the results obtained using the Burkert halo fit well with the observations and also match with the simulations (see Figure~\ref{fig:IRCs}, \ref{fig:SRCs}, \ref{fig:Mdyn-Mphot-Mbulge}, \ref{fig:fdm-out-obs}, \& \ref{fig:sims}). Therefore, in the rest of our paper, we solely discussed measurements obtained with the Burkert halo profile. We found the following:

\begin{itemize}
\item Our mass modeling approach has been proven to be robust to disentangle the various components of RCs, and it was able to determine the physical properties of baryons such as scale length, stellar and bulge mass, and DM structural properties such as core density and radius, as well as the fraction of DM (for details see Section~\ref{sec:results} and Appendix~\ref{sec:test-MM}). 

\item The DM fraction derived from the mass modeling outcomes (using  Equation~\ref{eq:fdm}) agrees very well with the results of the halo-model independent DM fraction (obtained in \citealt{GS21b}). Moreover, it is also similar to those predicted by different simulations: EAGLE, TNG100, and TNG50.

\item The results of the DM structural properties ($r_{_0}$ and $\rho_{_0}$) show that the DM cores at $z\sim 1$ are, on average, a factor of 0.3 dex smaller and 1.5 dex denser than their local counterparts. This suggests an expansion in the structure of DM halos over the last 6.5 Gyrs.

\item We also report that the expansion of DM cores is a function of circular velocity. The objects with a higher circular velocity expand their core more than those with low velocities. This suggests that circular velocity plays a crucial role in driving galaxy evolution.

\end{itemize}

We have shown in our previous work that the progenitors of local star-forming galaxies at $z\sim 1$ are similar in terms of total mass ($V_c^2(R) \propto M_{\mathrm{tot}}$); they are DM-dominated objects, and their stellar mass distribution evolves with time as a function of their circular velocity \citep[see,][]{GS21a, GS21b}. Now, this pilot study of dynamical mass modeling of high-$z$ RCs shows that DM distribution also changes over cosmic timescales, and it is a function of circular velocity. Therefore, we suggest that galactic processes that cause an evolution of the baryon distribution also have an impact on DM, and consequently DM distribution changes. In other words, DM responds to baryonic processes (e.g., AGN and supernova feedback) over the timescale of a few gigayears. If this work is confirmed by other observations, we will have the first observational evidence of "gravitational potential fluctuations" in the inner region of galaxies, originally theoretically predicted and reviewed by \citet[][and references therein]{Andrew2014}. This piece of evidence will have a strong implication for cross-checking simulations, modeling galaxy evolution, as well as determining the astrophysical and particle nature of DM.

\subsection*{Acknowledgments}
We thank the anonymous referee for their constructive comments and suggestions, which have significantly improved the quality of the manuscript. We thank K. Oman for providing us with the hydro-dynamical simulations as well as discussion and comments on the results. G.S. thanks J. Freundlich for passing the NIHAO simulations' concentration results. G.S. also thanks A. Marasco, M. Petac and S. Goswami for their fruitful discussion and various comments in the entire period of this work. GvdV acknowledges funding from the European Research Council (ERC) under the European Union’s Horizon 2020 research and innovation programme under grant agreement No 724857 (Consolidator GrantArcheoDyn).


%
\bibliographystyle{mnras} 
\bibliography{Gsharma2021.bib} 
%

\appendix

\section{Simulations}
\label{sec:sims-data}

\begin{figure*}[h]
  \begin{center}
   \includegraphics[angle=0,height=4.5truecm,width=18.5truecm]{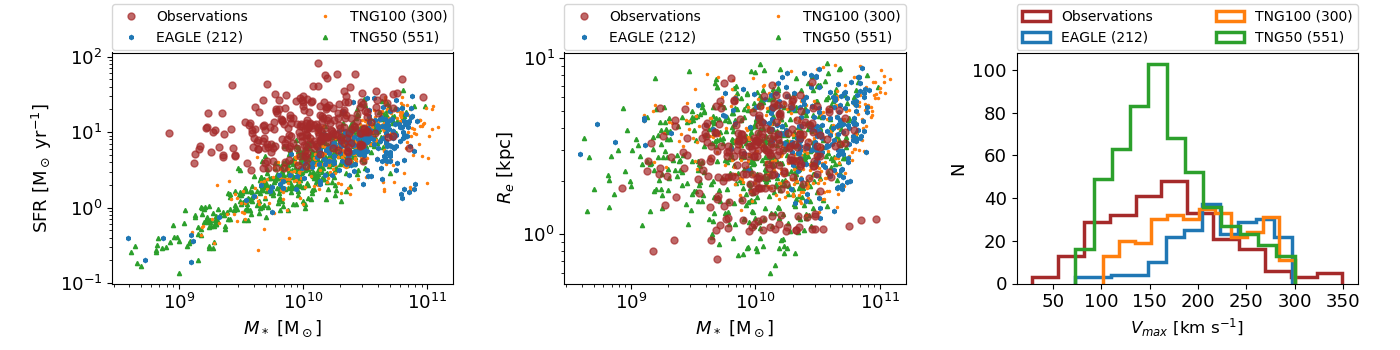}       
    
    \caption{Comparison of the physical parameters of the simulations (EAGLE, blue dots; TNG100, orange stars; TNG50, green triangles) with observations (brown filled circles). {\em Left panel:} Star formation rate as a function of stellar mass, the so-called main sequence of SFGs. {\em Middle panel:} Effective radii of galaxies as a function of stellar mass, the so-called mass-size relation. {\em Right panel:} Distribution of the maximum circular velocity of the simulated sample compared to $V_{out}=V_c$ of the observations.}
    \label{fig:sims-selections}
  \end{center}
\end{figure*}

\begin{figure*}[h]
  \begin{center}
   \includegraphics[angle=0,height=7.0truecm,width=9.0truecm]{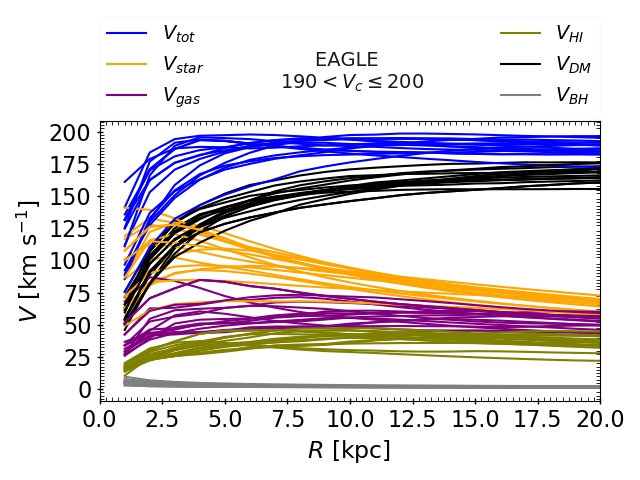}
   \includegraphics[angle=0,height=7.0truecm,width=9.0truecm]{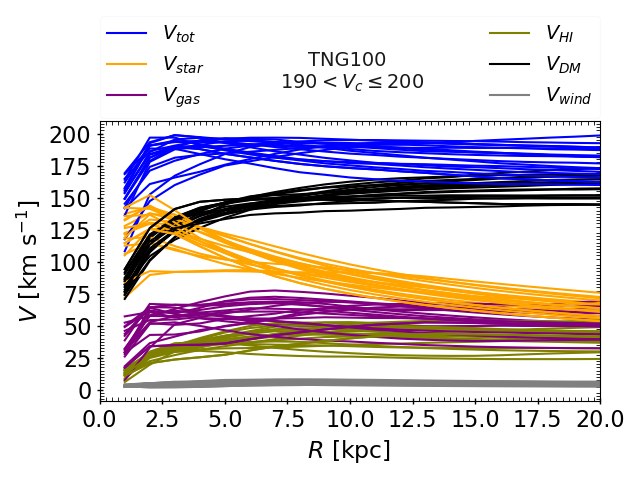}  
   \includegraphics[angle=0,height=7.0truecm,width=9.0truecm]{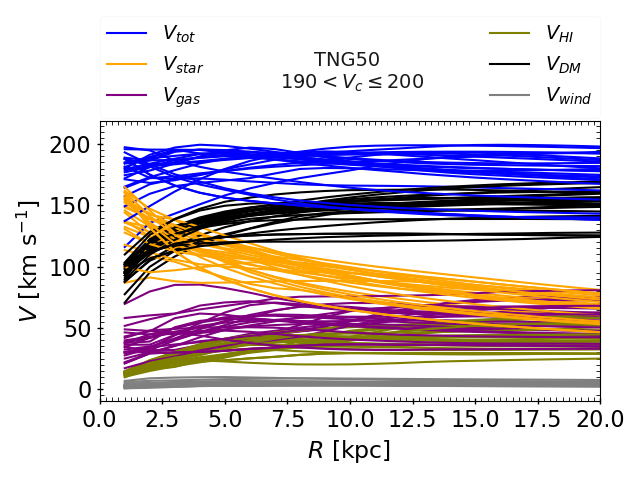}         
    
    \caption{Individual circular velocity curves of galaxy simulations from first-to-last EAGLE, TNG100, and TNG50, respectively. The color for each panel (and plot) is given in the legend as well as the name of the simulation, and the velocity bin is shown in the title of each plot.}
    \label{fig:sims-RCs}
  \end{center}
\end{figure*}

\begin{figure*}[h]
  \begin{center}
   \includegraphics[angle=0,height=7.0truecm,width=9.0truecm]{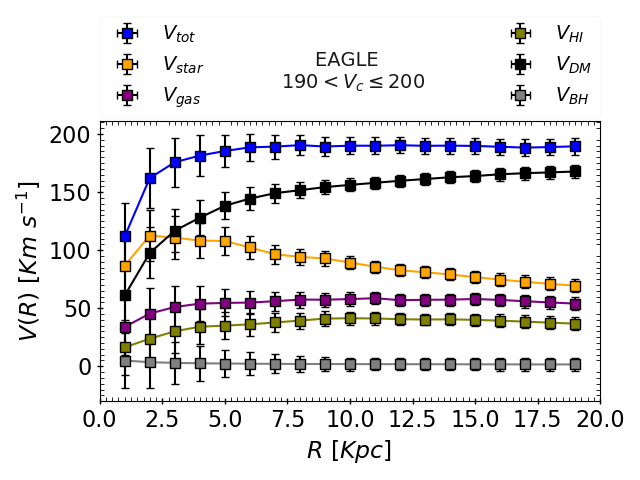}
   \includegraphics[angle=0,height=7.0truecm,width=9.0truecm]{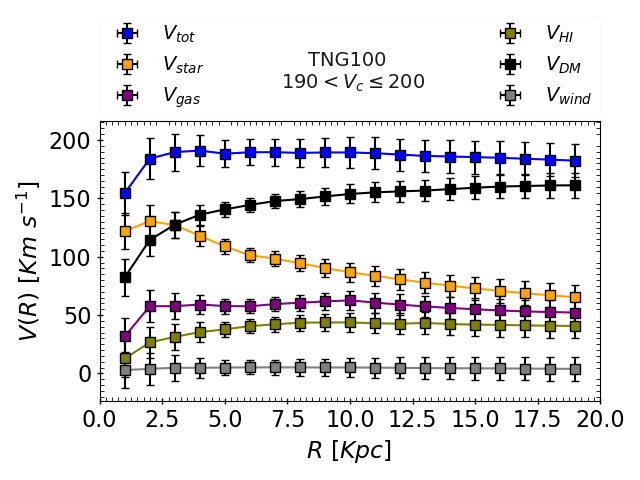}  
   \includegraphics[angle=0,height=7.0truecm,width=9.0truecm]{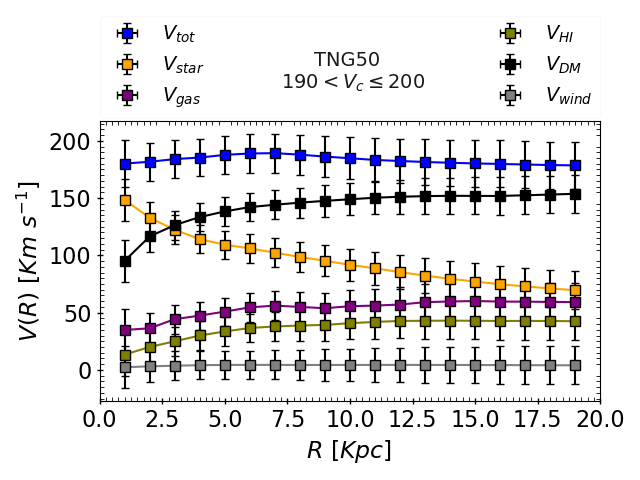}         
    
    \caption{Coadded and binned circular velocity curves of galaxy simulations from first to last EAGLE, TNG100, and TNG50. The color for each panel (and plot) is given in the legend as well as the name of the simulation, and the velocity bin is shown in the title of each plot.}
    \label{fig:sims-RCs}
  \end{center}
\end{figure*}

\begin{figure*}[h]
  \begin{center}
   \includegraphics[angle=0,height=6.5truecm,width=9.0truecm]{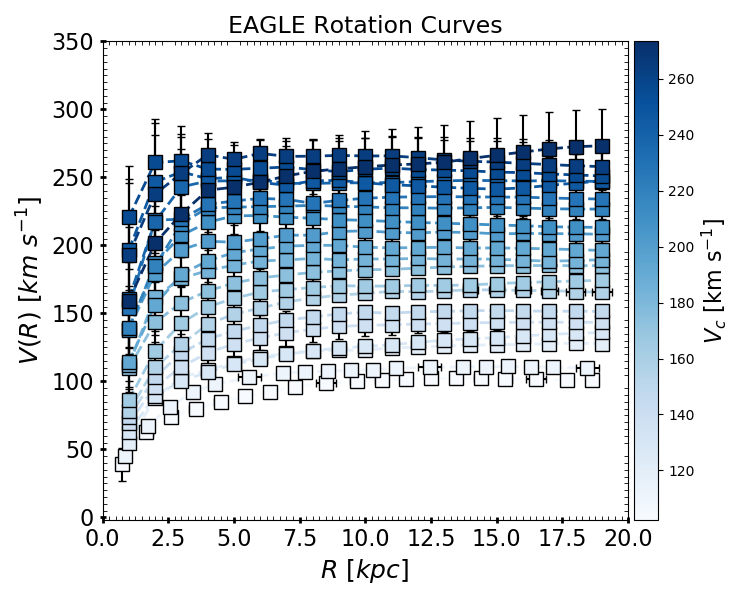}
   \includegraphics[angle=0,height=6.5truecm,width=9.0truecm]{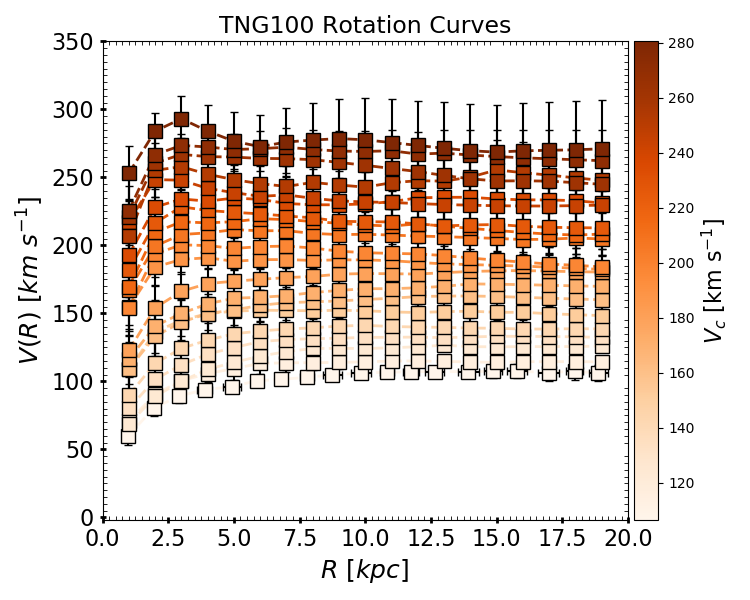}  
   \includegraphics[angle=0,height=6.5truecm,width=9.0truecm]{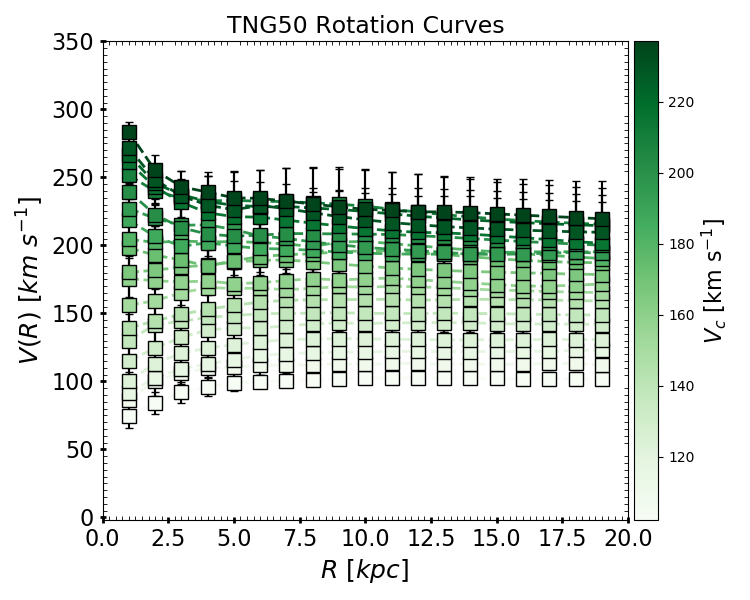}         
    
    \caption{Binned RCs of all velocity bins of galaxy simulations from first to last EAGLE, TNG100, and TNG50. The color maps give an idea of their increasing circular velocities.}
    \label{fig:sims-RCs}
  \end{center}
\end{figure*}

In addition to observations, we also analyzed recent state-of-the-art galaxy simulations, EAGLE \citep{EAGLE} and IllustrisTNG \citep{TNG}. For the purposes of this paper, we selected central galaxies from the simulations using a friends-of-friends halo finder approach \citep{Davis1985} and the SUBFIND algorithm \citep{Springel2001, Dolag2009}. Gas and star particles are associated with the galaxy of their nearest neighboring DM particle if they belong to one, and if they are gravitationally bound to the object. The center of a galaxy is defined as the location of its particle and cell (of any type) with the minimum gravitational potential energy evaluated over the particles belonging to the galaxy -- we take this as the origin of the profiles assumed in Section~\ref{sec:DM-profile}. 

To match the observed and simulated galaxies, we further filtered the central galaxies using the following selection criteria: $z= 0.85$, $M_{\mathrm{star}} = 10^8-10^{11.6} \ \mathrm{M_\odot}$, SFR = $0.1-100 \ \mathrm{M_\odot yr^{-1}}$, $R_e = 0.03-30 \ \mathrm{kpc}$, $V_c = 75- 300 \ \mathrm{km \ s^{-1}} $, and  $V_{rot}/\sigma >1$. Here, $V_c$ is the maximum circular velocity of the galaxies within the radius of 100 kpc. The $V_{rot}/\sigma$ is the ratio between the maximum rotation (azimuthal) velocity to the velocity dispersion. With this selection criteria, we obtained 212 EAGLE, 300 TNG100, and 551 TNG50 galaxies. We present this selected sample together with the observed data in Figure~\ref{fig:sims-selections}, where we show the main sequence of SFGs, the mass-size relation, and the velocity distribution. We can notice that our observed sample agrees very well with the simulations in terms of their effective radii, stellar masses, and circular velocities. However, the SFR of the simulated sample is lower by a factor of 0.37 dex, which is one of the well-known problems in current simulations \citep{Whitaker2014, Donnari2019}. In spite of this mismatch, we consider the selected simulated galaxy sample as an optimal sample for a detailed comparison with our data, as it is a representative for the population of typical star-forming rotation-dominated galaxies at $z\sim 1$ in the $\Lambda$CDM universe.

To compare the simulations with observations, we extracted the mass profile of simulated galaxies expressed as the circular velocity curve ($V_c(R) = \sqrt{GM/R}$) and its constituent components (i.e., stars, gas, and DM). We coadded the simulated circular velocity curves using a similar approach as used for observations, and binning was performed using the RMS statistic. Briefly, we divided our simulated galaxies into 20 velocity bins between 100-300 $\mathrm{km \ s^{-1}}$ with a step size of 10 $\mathrm{km \ s^{-1}}$. Then each bin was coadded and binned to obtain a single circular velocity curve. We would like to note that in the simulations, we know the exact total circular velocity of the galaxy, as well as its different components, namely stars ($V_{\mathrm{star}}$), gas ($V_{H2}$ and $ V_{\mathrm{HI}}$), and DM ($V_{\mathrm{DM}}$). Therefore, in the binning process, we binned the total circular velocity, as well as its various components. Moreover, we also binned the physical quantities associated with each galaxy, such as $R_{\mathrm{D}}$ and  $M_{\mathrm{star}}$. In Figure~\ref{fig:sims-RCs}, we show an example case of coadding and  binning, as well as the final coadded circular velocity curves from all three simulations, and in Table~\ref{tab:sims-params}, we tabulate all of their relevant physical quantities. These circular velocity curve data were then used in the investigation of the DM fraction (see Section~\ref{sec:results}). We remark that the TNG100 and TNG50 simulations are the same and only differ in terms of resolution. However, we still notice a small dissimilarity in the shape of their circular velocity curves. Nevertheless, in this work, we avoid making any interpretations on the circular velocity curves derived from the simulations, as it is beyond the scope of this work and will be covered in our future work for which we plan to make the synthetic datacubes of simulated galaxies and determine the RCs using the 3D Barolo, as suggested by \citet{Kyle2018} and \citet{Kyle2019}.

\section{Testing mass modeling approach}
\label{sec:test-MM}
\begin{figure*}
  \begin{center}
   \includegraphics[angle=0,height=5.5truecm,width=8.0truecm]{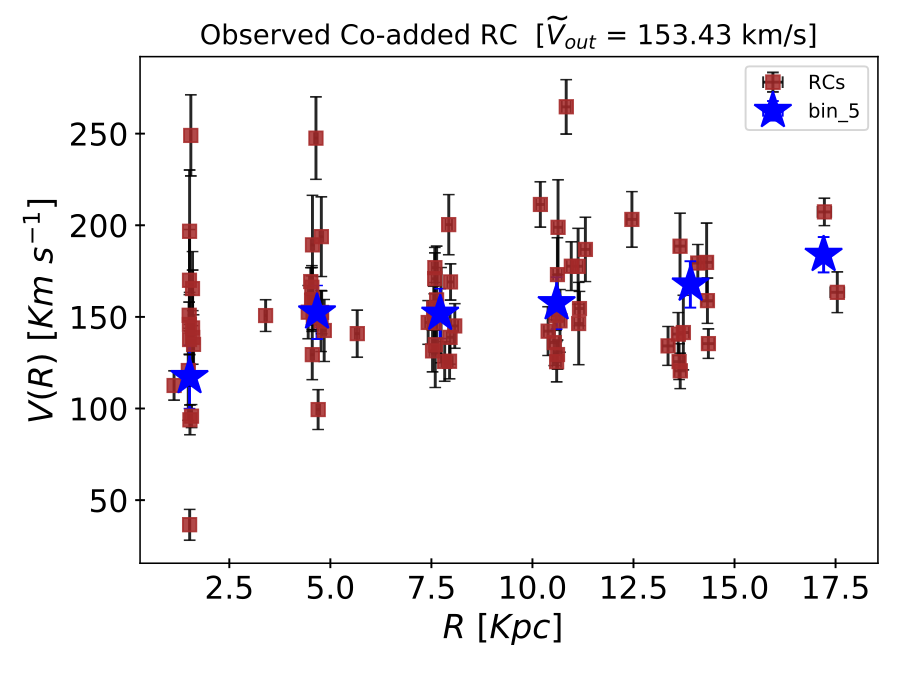}

   \includegraphics[angle=0,height=6.5truecm,width=8.0truecm]{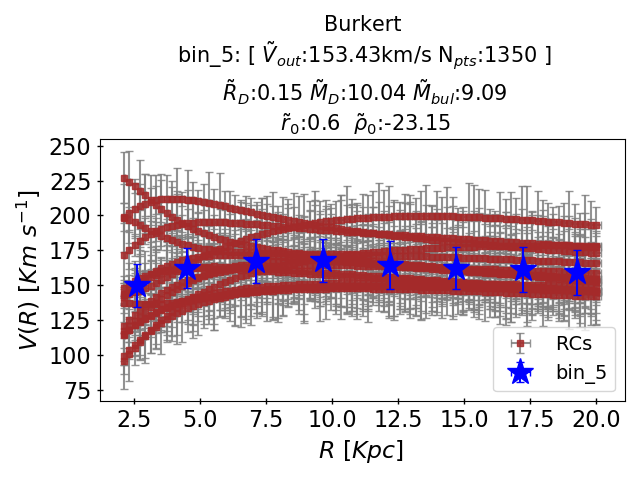}  
   \includegraphics[angle=0,height=6.5truecm,width=8.0truecm]{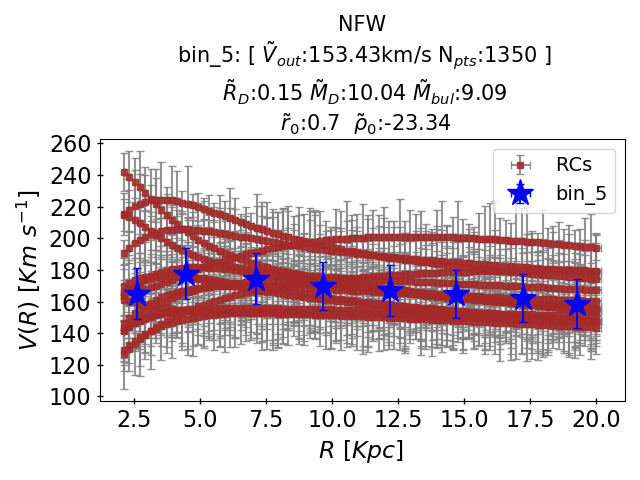}         
    
    \caption{Example of synthetic coadded RCs. {\em Top panel:} Observed coadded and binned RC, which is shown as a reference. {\em Bottom panel:} Coadded and binned synthetic RCs in the case of the Burkert and the NFW dark matter halo, which are on the left and right, respectively. The color code in all of the panels are the same and given as follows: the brown square represents the individual RCs and the blue stars represent the coadded RC.}
    \label{fig:SynCRCs}
  \end{center}
\end{figure*}

\begin{figure*}
  \begin{center}
   \includegraphics[angle=0,height=7.9truecm,width=8.5truecm]{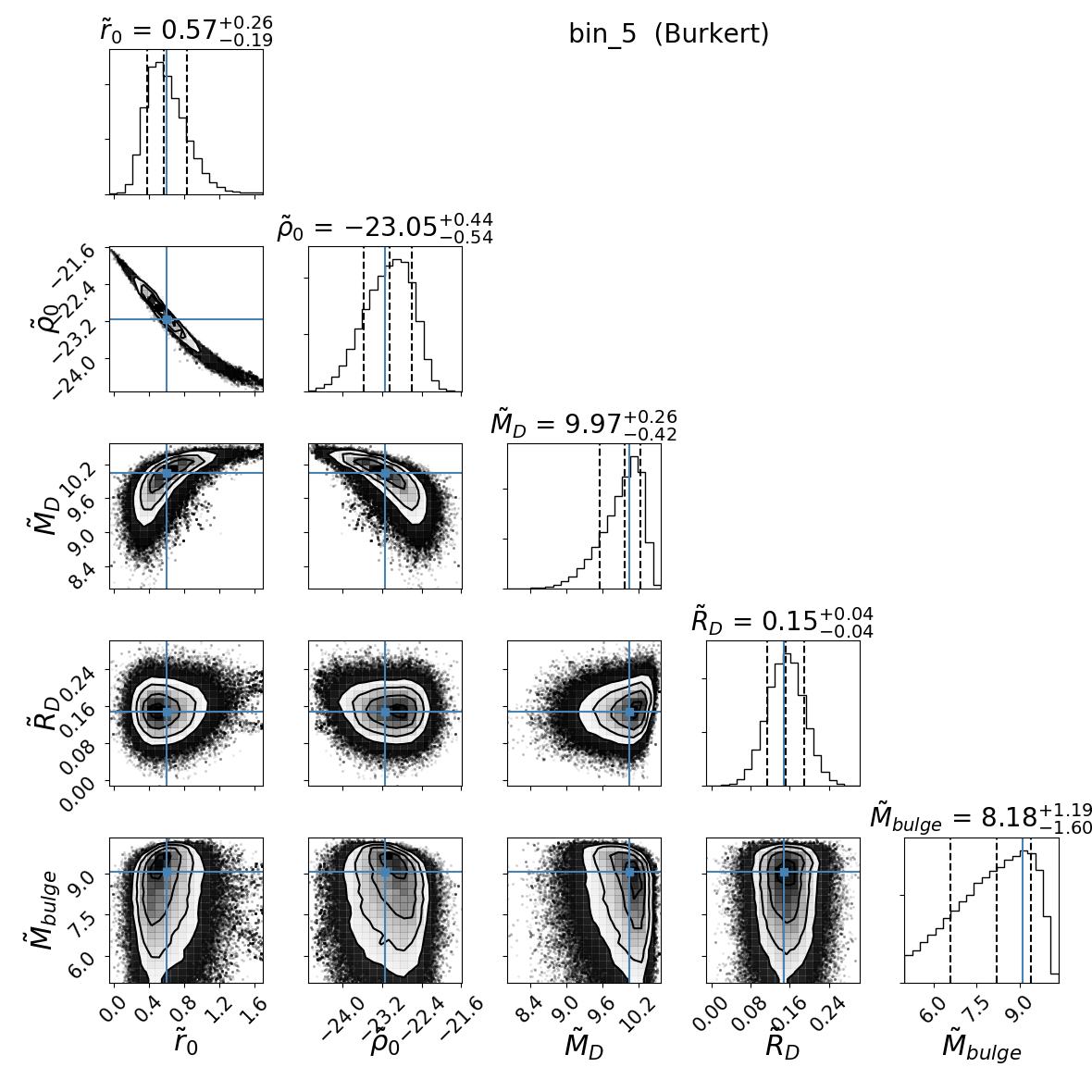}  
   \includegraphics[angle=0,height=7.9truecm,width=8.5truecm]{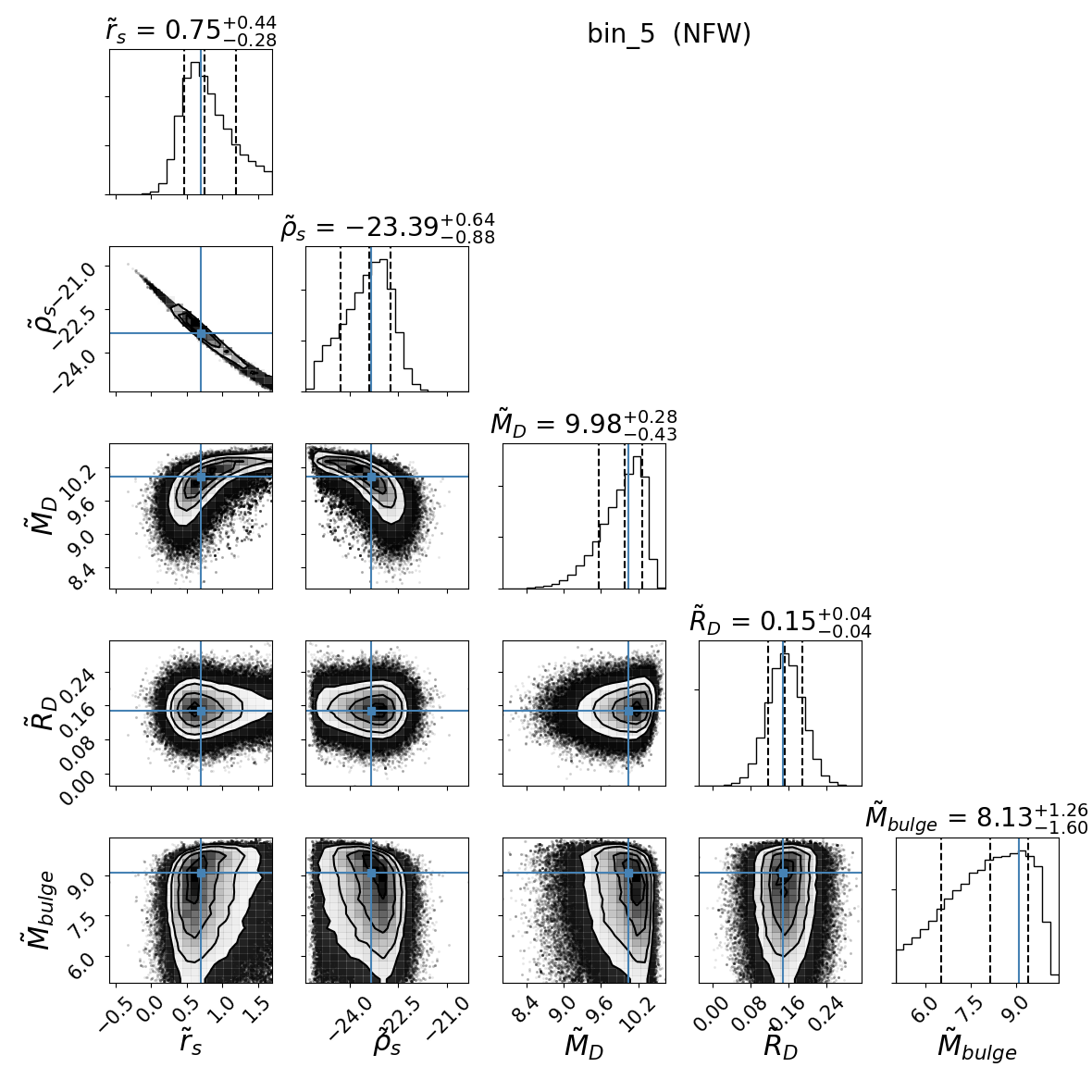}         
    
    \caption{Example of MCMC posteriors of mass-modeled synthetic coadded RCs, in the case of the Burkert and the NFW halo, on the left and right, respectively. Here, we show that the posteriors correspond to the synthetic coadded RC shown in Figure~\ref{fig:SynCRCs}. The best-fit parameters are printed above each posterior with a $1\sigma$ uncertainty. The vertical black dashed lines show the best-fit central value along with the $1\sigma$ upper and lower limit, and the true values are shown by blue crosses.}
    \label{fig:PSynCRCs}
  \end{center}
\end{figure*}

To test the performance of our mass modeling approach, we verified it on synthetic RCs. In this section, we explain this procedure in detail.

\subsection{Synthetic RCs}
To generate the synthetic RCs, we took the reverse approach, that is, we used the best-fitting parameters obtained from the mass modeling of the individual RCs, namely $\vec{\theta} = [M_D, R_D, M_{\mathrm{bulge}}, r_{0,s}, \rho_{0,s}]$, gaseous disk parameters = [$M_{H2}, M_{ HI }, R_{H2}, R_{ HI }$], and SFR. The redshift of the synthetic RCs was fixed to $z=0.85$. Then, the synthetic RCs were generated using the baryon and DM models defined in Section~\ref{sec:models}, and in the end we assigned to each RC the uniform error of 5-25 km/sec typically observed in the true RCs. In this way, we obtained 228 individual synthetic RCs, for which we know the exact baryon and DM content. An example of such individual RCs is shown in the bottom panel of Figure~\ref{fig:SynCRCs}. We note that we generated synthetic RCs for both the Burkert and NFW halo profiles. We used these synthetic RCs to test our mass modeling approach.

\subsection{Synthetic coadded RCs}
Coaddition of synthetic RCs was performed using the same strategy applied on observed RCs, defined in Section~\ref{sec:CRCs}. That is, each synthetic coadded RC contain 15 synthetic individual RCs; they were binned using weighted mean statistics. During the binning process, we also binned the parameters space $\vec{\theta}$ and gas disk parameters, yielding $\tilde{M}_D, \ \tilde{R}_D, \ \tilde{M}_{\mathrm{bulge}}, \ \tilde{r}_0, \ \tilde{\rho}_0, \  \tilde{M}_{H2}, \ \tilde{M}_{ HI }, \ \tilde{R}_{H2},$ and $ \ \tilde{R}_{ HI }$ for synthetic coadded RCs, which we used as a central values in mass modeling (if required). An example of a synthetic coadded RC is shown in Figure~\ref{fig:SynCRCs}.

\subsection{Mass modeling of synthetic coadded RCs}
For the mass modeling of synthetic coadded RCs, we used the same technique as applied on observed coadded RCs, and described in Section~\ref{sec:technique}. We examined the parameter space ($\vec{\theta}$) using the MCMC sampler. In the fitting process, we kept the Gaussian prior in the log scale for $R_D$ and $M_D$ with a 0.25 dex dispersion around their central values. For the rest of the parameters ($\vec{\theta}$), we used the flat prior in the log scale with ranges given in Table~\ref{tab:params}. The gas mass (atomic and molecular) depends on the scaling relations \citep{Tacconi2018, Lagos2011} and it was derived in each MCMC run using the running value of the stellar mass, the fixed SFR, and $z=0.85$. That is, the best-fit gas mass depends on the best-fit stellar mass, SFR, and $z$. The scale lengths of the gaseous disks ($R_{H2}$ and $ R_{HI}$) are fixed.
\\
\\
{\bf Results:} In the lower panel of Figure~\ref{fig:SynCRCs}, we show an example of synthetic individual and coadded RCs for bin number five (bin\_5) in the case of Burkert and NFW halo profiles, which are on the left and right, respectively. These synthetic RCs are shown in comparison with ther observed coadded RCs (upper panel) of the same velocity bin. The binned values (i.e., the true values) corresponding to each coadded RC are given in the title of each RC plot. We allowed mass modeling of these RCs using the  aforementioned procedure. In Figure~\ref{fig:PSynCRCs}, we show the posterior of bin\_5; as one can notice, our mass modeling is capable of retrieving the true $\vec{\theta}$ space within a $1\sigma$ uncertainty. For all velocity bins tested (but not shown here), similarly good results were found. Therefore, we conclude that our mass modeling approach has proven robust to disentangle the various components of RCs.

\section{Tables}
\label{sec:table}
In Table~\ref{tab:obs-params}, we provide the physical parameters of observed galaxies (in the case of coadded RCs). In particular, we provide relevant details on coadded RCs (binning), averaged physical quantities ($\tilde{R}_{\mathrm{D}}$, $\tilde{V}_c$, $\tilde{M}_*$, $\mathrm{\tilde{SFR}}$, and $\tilde{R}_{\mathrm{H2}}$) per bin, and best-fit modeled parameter space $\vec{\theta}$ and the DM fraction, in the case of the Burkert and NFW halo. In Table~\ref{tab:sims-params}, we provide the relevant physical parameters ($\tilde{R}_{\mathrm{D}}$, $\tilde{V}_c$, $\tilde{M}_*$, $\tilde{R}_{gas}$, and $\tilde{f}_{\mathrm{\tiny{DM}}}$) of galaxy simulations: EAGLE, TNG100, and TNG50.

\section{Extra Figures}
\label{sec:extra-figs}
\begin{figure*}
 \begin{center}
  \includegraphics[angle=0,height=9.0truecm,width=9.0truecm]{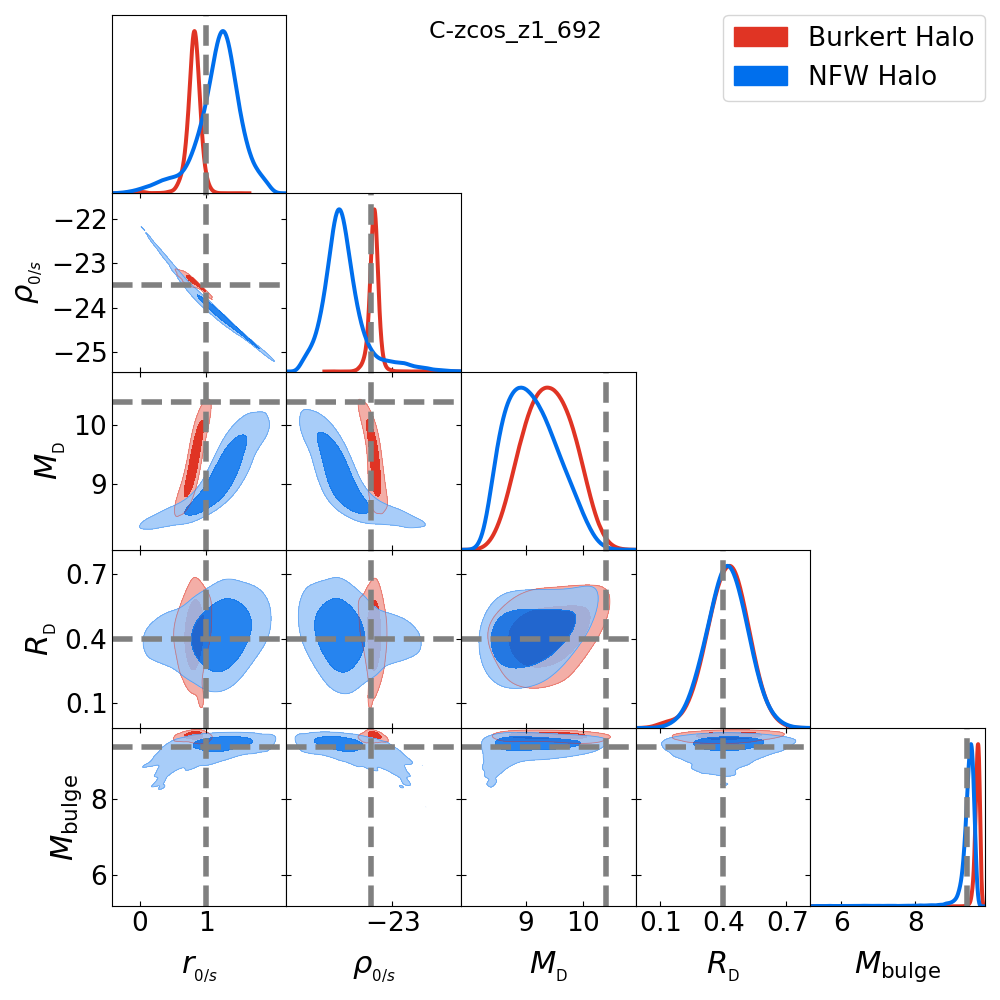} 
  \includegraphics[angle=0,height=9.0truecm,width=9.0truecm]{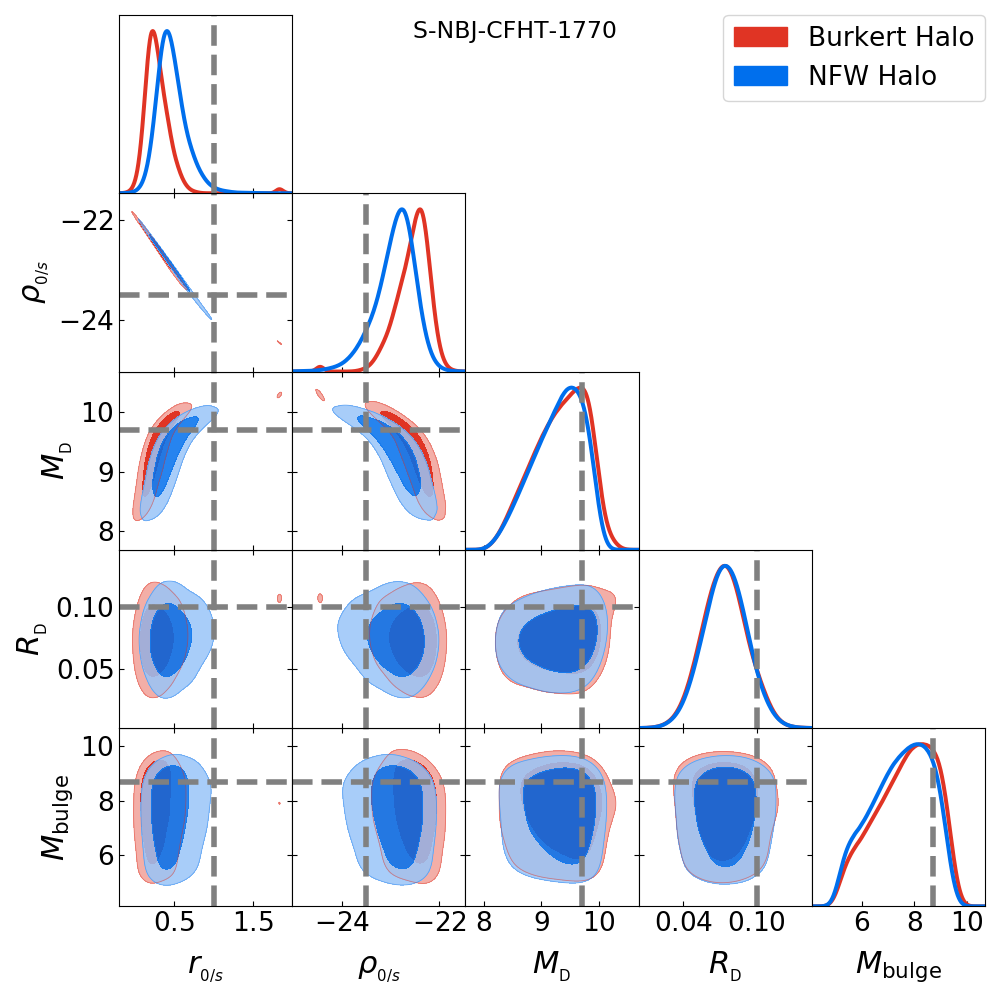} 

    \caption{\underline{Individual RC mass modeling:} A few examples of full posterior distributions (MCMC output) of the estimated parameters for the Burkert (red) and NFW (blue) halo. The vertical and horizontal gray dashed line on the posterior plots show the initial guess (or mean value in the case of a Gaussian prior) of parameter space $\vec{\theta}$. This figure is an extended version of Figure~\ref{fig:IRCs}. }
    \label{fig:IRCs_full}
  \end{center}
\end{figure*}

\begin{figure*}
 \begin{center}
  \includegraphics[angle=0,height=9.0truecm,width=9.0truecm]{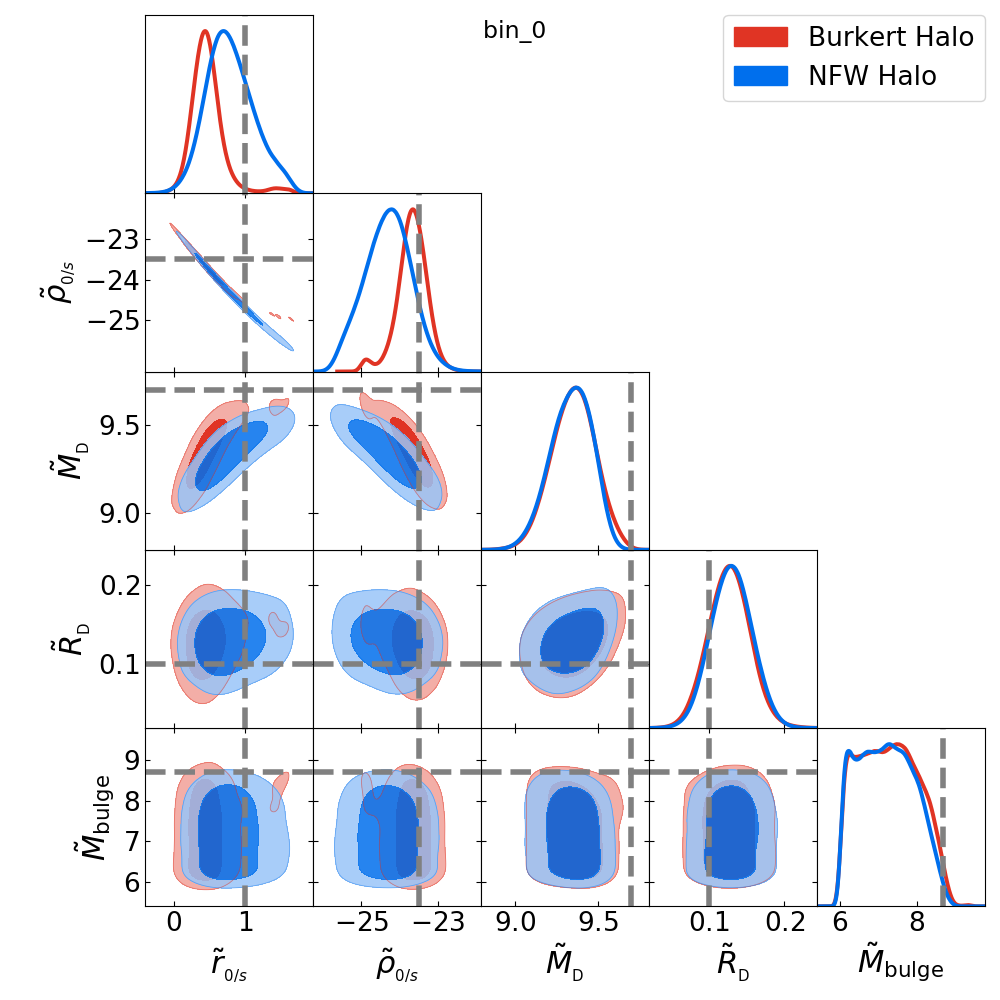} 
    \includegraphics[angle=0,height=9.0truecm,width=9.0truecm]{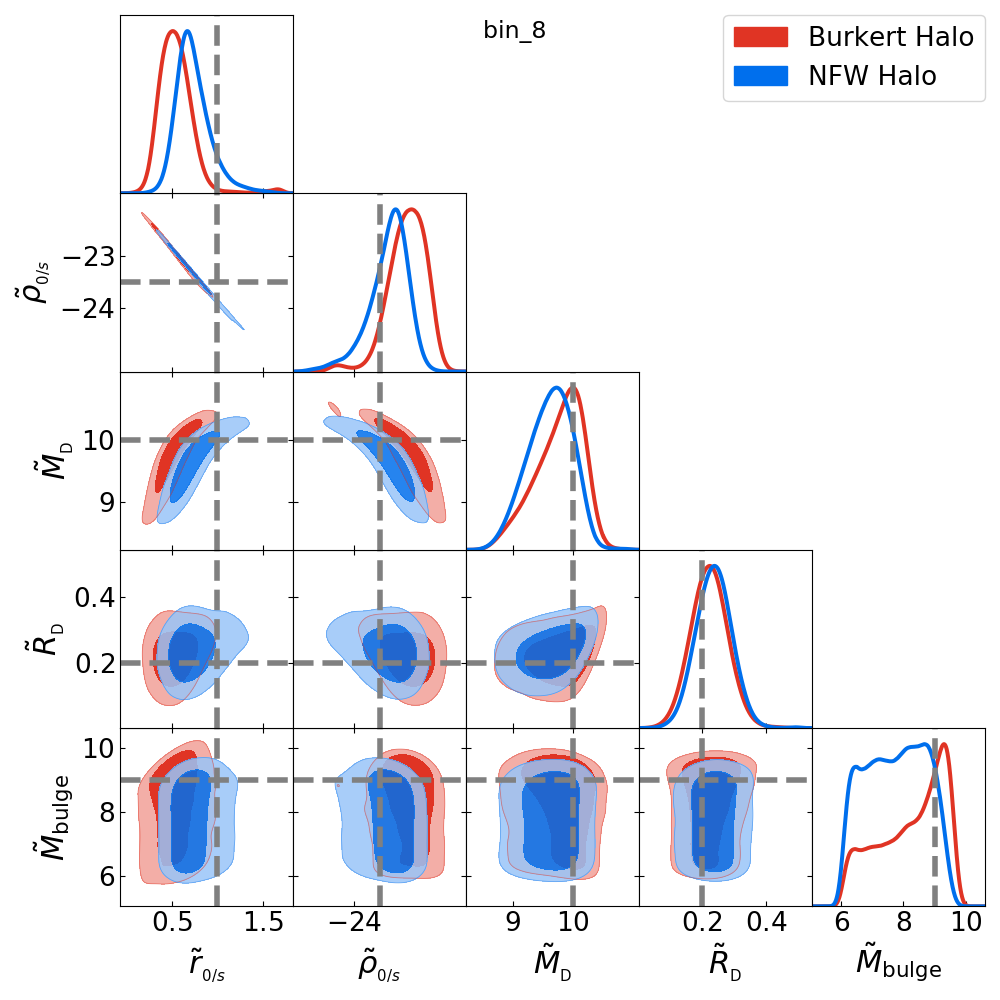}  

    \caption{{\underline{Coadded RC mass modeling:} A few examples of full posterior distributions (MCMC output) of the estimated parameters for the Burkert (red) and NFW (blue) halo. The vertical and horizontal gray dashed line on the posterior plots show the initial guess (or mean value in the case of the Gaussian prior) of parameter space $\vec{\theta}$. This figure is an extended version of Figure~\ref{fig:SRCs}.}}
    \label{fig:SRCs_full}
  \end{center}
\end{figure*}

In Figures~\ref{fig:IRCs_full} \& \ref{fig:SRCs_full}, we provide the posterior  distribution of full parameter space  $\vec{\theta}$ for individual and coadded RCs, respectively. 

\begin{table*}
\tiny
\begin{filecontents*}{Obserbations_parameters.csv}
Bin,RC_pts,Rd,Vc,Md,SFR,RH2,Brd,Bmd,Bmbu,Brc,Brho0,Nrd,Nmd,Nmbu,Nrc,Nrho0,Bfdm,Nfdm
0,15 (64),0.11,1.92,9.69,0.92,1.02,0.13,9.35,7.24,0.44,7.52,0.13,9.33,7.23,0.75,6.92,0.45,0.44
1,15 (70),0.27,1.98,9.86,0.87,1.16,0.26,9.64,8.86,0.63,7.47,0.25,9.3,8.33,0.43,7.96,0.59,0.7
2,15 (65),0.22,2.06,9.75,1.02,0.93,0.22,9.3,9.03,0.34,7.92,0.22,9.24,8.56,0.25,8.21,0.4,0.44
3,15 (68),0.29,2.1,9.83,1.07,1,0.25,10.03,8.99,1.17,6.93,0.28,8.8,8.46,0.23,8.5,0.49,0.52
4,15 (67),0.19,2.14,9.89,0.98,1.11,0.19,9.96,8.48,0.65,7.79,0.2,9.46,7.97,0.63,7.88,0.67,0.8
5,15 (74),0.15,2.19,10.09,1.12,1.11,0.15,9.86,8.05,0.6,8.02,0.16,9.46,7.76,0.7,7.83,0.76,0.83
6,15 (78),0.26,2.2,10.12,0.99,1.03,0.28,10.3,8.05,1.07,7.14,0.26,9.96,7.76,1.07,7.09,0.51,0.71
7,15 (69),0.27,2.22,10.03,1.16,1.1,0.25,10.18,8.49,0.85,7.61,0.27,9.59,7.82,0.83,7.66,0.66,0.81
8,15 (70),0.21,2.24,10.02,1.02,1.04,0.23,9.85,8.43,0.52,8.25,0.24,9.66,7.78,0.71,7.89,0.79,0.83
9,15 (86),0.25,2.27,10.23,1.05,1.26,0.23,10.19,9.29,0.88,7.68,0.24,9.55,9.02,0.98,7.49,0.7,0.88
10,15 (68),0.16,2.29,10.33,1.08,1.05,0.15,10.3,8.92,0.78,7.79,0.15,9.96,8.59,0.81,7.77,0.58,0.76
11,15 (72),0.24,2.31,10.23,1.04,1,0.29,10.21,7.73,0.52,8.38,0.35,10.42,7.65,0.96,7.47,0.77,0.68
12,15 (69),0.28,2.34,10.03,1.3,1.02,0.3,10.54,8.69,0.88,7.89,0.3,10.08,8.03,0.92,7.81,0.7,0.85
13,15 (65),0.21,2.41,10.35,0.91,1.09,0.22,10.44,8.03,0.69,8.25,0.22,10.04,7.78,0.88,7.96,0.75,0.89
14,15 (83),0.33,2.44,10.25,0.99,0.99,0.31,10.78,9.19,1.09,7.79,0.29,10.35,9.32,1.22,7.5,0.66,0.84
15,3 (13),0.36,2.53,10.4,1.4,1.14,0.32,10.94,10.12,0.99,7.93,0.33,10.6,10.09,1.08,7.74,0.59,0.78
\end{filecontents*}
\pgfplotstabletypeset[
    col sep=comma,
    string type,
    every head row/.style={%
        before row={
       		 \hline
			 \multicolumn{19}{|c|}{Physical Parameters: OBSERVATIONS} \\
			\hline
				\multicolumn{2}{c}{\underline{Binning Details}} & \multicolumn{5}{|c||}{\underline{Photometric measurements}} & \multicolumn{5}{|c|}{\underline{$\vec{\theta}$: Burkert Halo}} & \multicolumn{5}{|c|}{\underline{$\vec{\theta}$: NFW Halo}} & \multicolumn{2}{|c}{\underline{DM Fraction}}  \\[0.7ex]
        },
        after row=\hline
    },
    every last row/.style={after row=\hline},
    columns/Bin/.style={column name=No., column type=l},    
    columns/RC_pts/.style={column name=Dpts, column type=l|},
    columns/Rd/.style={column name=$\tilde{R}_{\mathrm{D}}$, column type=l},
    columns/Vc/.style={column name=$\tilde{V}_c$, column type=l},   
    columns/Md/.style={column name=$\tilde{M}_*$, column type=l},
    columns/SFR/.style={column name=$\mathrm{\tilde{SFR}}$, column type=l},
    columns/RH2/.style={column name=$\tilde{R}_{\mathrm{H2}}$, column type=l|},      
    columns/Brd/.style={column name=$\tilde{R}_{\mathrm{D}}$, column type=l},
    columns/Bmd/.style={column name=$\tilde{M}_{\mathrm{D}}$, column type=l},
    columns/Bmbu/.style={column name=$\tilde{M}_{bul}$, column type=l},       
    columns/Brc/.style={column name=$\tilde{r}_0$, column type=l},
    columns/Brho0/.style={column name=$\tilde{\rho}_0$, column type=l|},
    columns/Nrd/.style={column name=$\tilde{R}_{\mathrm{D}}$, column type=l},          
    columns/Nmd/.style={column name=$\tilde{M}_{\mathrm{D}}$, column type=l},
    columns/Nmbu/.style={column name=$\tilde{M}_{bul}$, column type=l},
    columns/Nrc/.style={column name=$\tilde{r}_0$, column type=l},
    columns/Nrho0/.style={column name=$\tilde{\rho}_0$, column type=l|},
    columns/Bfdm/.style={column name=$f_{\mathrm{\tiny{DM}}}^{\mathrm{\tiny{BUR}}}$, column type=l},
    columns/Nfdm/.style={column name=$f_{\mathrm{\tiny{DM}}}^{\mathrm{\tiny{NFW}}}$, column type=l}, 
    ]{Obserbations_parameters.csv}

\caption{Structural (and physical) parameters of the observed galaxies derived from their coadded RCs. {\em Columns 1-2:} Bin number and total number of RCs, respectively. For example, the notation 15(67) indicates that we have coadded 15 RCs and that they contain 67 data-points. {\em Columns 3-7:} Photometric measurements of physical quantities $\tilde{R}_{\mathrm{D}}$, $\tilde{V}_c$, $\tilde{M}_{\mathrm{D}}$, $\mathrm{\tilde{SFR}}$, and $\tilde{R}_{\mathrm{H2}}$, respectively. The tilde $\tilde{•}$ on top of each quantity means that it is a statistically measured quantity. {\em Columns 8-12:} Modeled physical parameters ($\tilde{R}_{\mathrm{D}}$, $\tilde{M}_{\mathrm{D}}$, $\tilde{M}_{\mathrm{bul}}$, $\tilde{r}_{_0}$, and $\tilde{\rho}_{_0}$, respectively) for baryons and DM in the Burkert halo case. {\em Columns 13-17:} Modeled physical parameters ($\tilde{R}_{\mathrm{D}}$, $\tilde{M}_{\mathrm{D}}$, $\tilde{M}_{\mathrm{bul}}$, $\tilde{r}_{_s}$, and $\tilde{\rho}_{_s}$, respectively) for baryons and DM in the NFW halo case. {\em Columns 18-19:} \rm{DM fraction within $R_{out}$} in the case of the Burkert and NFW halo, respectively. Here, we provide each quantity in $\log$ (except the DM fraction) and their physical units are as follows: $\tilde{R}_{\mathrm{D}} \ [\mathrm{kpc}]$, $\tilde{V}_c \ [\mathrm{km \ s^{-1}}]$, $\tilde{M}_{\mathrm{D}} \ [\mathrm{M_\odot}]$, $\mathrm{\tilde{SFR}} \ [\mathrm{M_\odot \ yr^{-1}}]$, $\tilde{R}_{\mathrm{H2}} \ [\mathrm{kpc}]$, $\tilde{M}_{\mathrm{bul}} \ [\mathrm{M_\odot}]$, $\tilde{r}_{_0} \ [\mathrm{kpc}]$, and $\tilde{\rho}_{_0} \ [\mathrm{M_\odot \ kpc^{-3}}]$. }
\label{tab:obs-params}
\end{table*}

\begin{table*}
\tiny
\begin{filecontents*}{Simulations_parameters_fdm.csv}
bin,EN,T100N,T50N,Erd,T100rd,T50rd,Evc,T100vc,T50vc,Emd,T100md,T50md,Ergas,T100rgas,T50rgas,Efdm,T100fdm,T50fdm
0,1,4,38,0.43,0.34,0.29,2.01,2.03,2.01,9.1,9.15,9.28,0.97,1.37,1.45,0.88,0.78,0.74
1,1,11,29,0.35,0.37,0.33,2.04,2.06,2.06,9.32,9.39,9.51,1.23,1.56,1.49,0.82,0.74,0.74
2,3,9,35,0.43,0.41,0.36,2.1,2.1,2.1,9.61,9.53,9.66,1.54,1.55,1.5,0.81,0.76,0.7
3,4,15,53,0.45,0.43,0.37,2.12,2.12,2.12,9.74,9.69,9.8,1.5,1.58,1.51,0.8,0.74,0.7
4,4,11,37,0.51,0.36,0.39,2.16,2.15,2.16,9.81,9.84,9.94,1.62,1.71,1.58,0.81,0.72,0.68
5,7,24,64,0.44,0.36,0.36,2.18,2.19,2.18,9.91,9.95,10.01,1.72,1.69,1.61,0.81,0.71,0.68
6,5,12,38,0.46,0.41,0.39,2.22,2.21,2.21,10.07,10.09,10.11,1.63,1.74,1.61,0.76,0.69,0.66
7,16,16,40,0.46,0.42,0.34,2.24,2.23,2.23,10.17,10.16,10.19,1.73,1.78,1.65,0.76,0.71,0.65
8,8,22,31,0.45,0.42,0.33,2.27,2.26,2.25,10.29,10.26,10.29,1.83,1.77,1.68,0.76,0.66,0.66
9,15,19,25,0.45,0.32,0.37,2.28,2.28,2.28,10.35,10.3,10.34,1.83,1.77,1.71,0.7,0.65,0.65
10,14,22,25,0.46,0.32,0.43,2.3,2.3,2.29,10.41,10.36,10.4,1.82,1.82,1.68,0.73,0.63,0.65
11,25,16,21,0.44,0.3,0.36,2.33,2.33,2.31,10.49,10.47,10.44,1.89,1.84,1.72,0.7,0.63,0.63
12,12,17,12,0.34,0.37,0.26,2.34,2.34,2.3,10.5,10.5,10.42,1.99,1.89,1.74,0.66,0.62,0.6
13,12,22,21,0.44,0.37,0.34,2.36,2.35,2.32,10.61,10.54,10.47,2,1.86,1.79,0.69,0.61,0.62
14,13,14,12,0.38,0.41,0.34,2.37,2.38,2.34,10.66,10.65,10.53,2.05,1.92,1.76,0.66,0.65,0.62
15,18,14,12,0.39,0.38,0.3,2.39,2.38,2.36,10.71,10.67,10.65,2.07,1.9,1.77,0.67,0.63,0.57
16,19,15,11,0.39,0.45,0.16,2.39,2.41,2.35,10.7,10.81,10.5,2.08,2.01,1.78,0.68,0.63,0.57
17,12,19,8,0.36,0.4,0.39,2.41,2.41,2.37,10.75,10.84,10.59,2.09,1.97,1.8,0.66,0.59,0.63
18,14,13,1,0.39,0.51,0.32,2.43,2.43,2.34,10.8,10.93,10.65,2.15,2.01,1.71,0.67,0.64,0.61
19,7,5,12,0.59,0.5,0.21,2.44,2.45,2.37,10.83,10.99,10.68,2.17,2.1,1.85,0.77,0.64,0.52
\end{filecontents*}
\pgfplotstabletypeset[
    col sep=comma,
    string type,
    every head row/.style={%
        before row={
       		 \hline
			\multicolumn{19}{|c|}{Physical Parameters: SIMULATIONS} \\
			\hline
            \multicolumn{1}{c|}{Bin}& \multicolumn{3}{c|}{\underline{Number of RCs}}&  \multicolumn{3}{|c|}{\underline{$\tilde{R}_{\mathrm{D}}$}}&  \multicolumn{3}{|c|}{\underline{$\tilde{V}_c$}}&  \multicolumn{3}{|c|}{\underline{$\tilde{M}_*$}}&  \multicolumn{3}{|c|}{\underline{$\tilde{R}_{gas}$}}& \multicolumn{3}{|c}{\underline{$\tilde{f}_{\mathrm{\tiny{DM}}}$}}  \\
        },
        after row=\hline
    },
    every last row/.style={after row=\hline},
    columns/bin/.style={column name=No., column type=l|},    
    columns/EN/.style={column name=EAG, column type=l},
    columns/T100N/.style={column name=T100, column type=l},
    columns/T50N/.style={column name=T50, column type=l|},   
    columns/Erd/.style={column name=EAG, column type=l},
    columns/T100rd/.style={column name=T100, column type=l},
    columns/T50rd/.style={column name=T50, column type=l|},      
    columns/Evc/.style={column name=EAG, column type=l},
    columns/T100vc/.style={column name=T100, column type=l},
    columns/T50vc/.style={column name=T50, column type=l|},       
    columns/Emd/.style={column name=EAG, column type=l},
    columns/T100md/.style={column name=T100, column type=l},
    columns/T50md/.style={column name=T50, column type=l|},          
    columns/Ergas/.style={column name=EAG, column type=l},
    columns/T100rgas/.style={column name=T100, column type=l},
    columns/T50rgas/.style={column name=T50, column type=l|},
    columns/Efdm/.style={column name=EAG, column type=l},
    columns/T100fdm/.style={column name=T100, column type=l},
    columns/T50fdm/.style={column name=T50, column type=l}, 
    ]{Simulations_parameters_fdm.csv}

\caption{Physical parameters of galaxies in simulations (EAGLE, EAG; TNG100, T100; and TNG50, T50). {\em Column 1:} Bin number. {\em Columns 2-4:} Number of rotation curves in each bin. {\em Columns 5-7:} Disk radius. {\em Columns 8-10:} Circular velocity computed at $R_{\mathrm{out}}$. {\em Columns 11-13:} Stellar mass (includes bulge). {\em Columns 14-16:} Gas disk length. {\em Columns 17-19:} \rm{DM fraction within $R_{out}$}. Here quantities are provided in $\log$ (except the DM fraction) and their physical units are similar to the ones given in Table~\ref{tab:obs-params}.}
\label{tab:sims-params}
\end{table*}

\label{lastpage}

\end{document}